\documentclass[fleqn,aps,preprint,floatfix]{revtex4}
\usepackage{epsfig}

\font\fiverm=cmr5
\def \be {\begin{equation}}
\def \ee {\end{equation}}

\def \oh {\frac{1}{2}}
\def \tr {\hbox{tr}\,}
\def \one {{\bf 1}}
\def \avrg#1 {\langle #1 \rangle}
\def\Em{E_{M}}
\def\BE{\begin{equation}}
\def\EE{\end{equation}}
\def\BEA{\begin{eqnarray}}
\def\EEA{\end{eqnarray}}
\def\BC{\begin{center}}
\def\EC{\end{center}}
\def\re{{\rm Re}}
\def\Em{E_{M}}

\begin{document}

%
\preprint{DUKE-TH-92-40}

%
%

\title{Hamiltonian Dynamics\\ of Yang-Mills Fields on a Lattice}



%
\author{T.S. Bir\'o}
\address{Institut f\"ur Theoretische Physik, Justus-Liebig-Universit\"at,
D-6300 Giessen, Germany}
\author{C. Gong and B. M\"uller}
\address{Department of Physics, Duke University, Durham, NC 27708}
\author{A. Trayanov}
\address{NCSC, Research Triangle Park, NC 27709}
%
%

\date{\today}
\begin{abstract}
We review recent results from studies of the dynamics of classical
Yang-Mills fields on a lattice. We discuss the numerical techniques
employed in solving the classical lattice Yang-Mills equations in real
time, and present results exhibiting the universal chaotic behavior
of nonabelian gauge theories. The complete spectrum of Lyapunov exponents 
is determined for the gauge group SU(2).
We survey results obtained for the SU(3) gauge theory and other
nonlinear field theories. We also discuss the relevance of these
results to the problem of thermalization in gauge theories.
\end{abstract}
\pacs{PACS numbers: 14.60.Cd, 11.30.Er}
\maketitle

%
\baselineskip 6.333mm

\section{Introduction}

Knowledge of the microscopic mechanisms responsible for the local
equilibration of energy and momentum
carried by nonabelian gauge fields is important
for our understanding of non-equilibrium processes occurring in
the very early universe and in relativistic nuclear collisions.
Prime examples for such processes are baryogenesis during the
electroweak phase transition, the
creation of primordial fluctuations in the density of galaxies
in cosmology, and
the formation of a quark-gluon plasma in heavy-ion collisions.

Whereas transport and equilibration processes have been extensively
investigated in the framework of perturbative quantum field theory,
rigorous non-perturbative studies of nonabelian gauge theories have
been limited to systems at thermal equilibrium. We here review recent
numerical studies of real-time evolution in the classical
limit of lattice gauge theories. We will demonstrate that 
such an analysis can provide valuable insight into
the dynamical properties of nonabelian gauge theories 
at high excitation energies.  

This paper is organized as follows.
In section 2, we briefly discuss why a classical consideration
of gauge fields is relevant, and we review the early studies of evidence 
for chaotic behavior of Yang-Mills fields.
Section 3 is devoted to the results of our lattice
study on SU(2) gauge theory. After some general considerations, 
we introduce the lattice formalism of gauge fields, and 
we describe the numerical techniques used to solve the field equations. 
Then we show the exponential divergence of two nearby
trajectories, from which we can
conclude that the system is chaotic, and extract the maximal 
Lyapunov exponent. In the last two subsections of section 3,
we show how one can obtain the whole Lyapunov spectrum. 
After the experience with SU(2), results of several other theories are 
reviewed in section 4. These include compact U(1) and
SU(3) gauge theory, massless Higgs fields, massive SU(2) vector
fields, and finally the coupled SU(2) gauge-Higgs system.
In section 5, we discuss one application of chaoticity in nonabelian gauge 
theory, namely, the thermalization process of highly excited gauge fields.
We estimate the time needed to thermalize the system from the
Kolmogorov-Sinai entropy of the gauge field. This time agrees with
the results from thermal perturbation theory, supporting the view
that the thermalization of the long wavelength modes
is basically a classical process. Finally, in section 6, we
point out some possible avenues of future work in this field, 
both in regard to the method itself and to physical applications.

\section{Gauge Fields in the Classical Limit}

\subsection{General Considerations}

Let us start by discussing the classical limit of a simple 
system with a single degree of freedom which is described by a Hamiltonian
\be
H(p,x)=\frac{1}{2}p^{2} + V(x),
\ee
where $V(x)$ is some well defined potential. The quantum evolution
is defined
either by operator equations for $p$ and $x$ in Heisenberg picture,
\BEA
\dot{x}&=&p, \nonumber \\
\dot{p} &=& \partial_{x}V(x),
\EEA
or by a Schr\"odinger equation for the wave function,
\be
i\hbar\partial_{t} \Phi = {\hat H} \Phi.
\ee
Both are appropriate to describe
the behavior of a pure state. To admit
mixed states it is necessary to use the quantum Liouville equation
\be
i\hbar\partial_{t}\rho = [H,\rho]
\ee
for the density matrix $\rho$. 
For a system in a pure state, $\rho=|\Phi\rangle\langle\Phi|$,
the Liouville equation is reduced to the Schr\"odinger equation.

Quantum mechanics and classical mechanics are related by the
correspondence principle, which states that at large
quantum numbers or small $\hbar$ quantum mechanics has
classical mechanics as a limit. But it is not trivial  to actually construct
the correspondence, especially 
when the system considered is chaotic in the classical limit.
Quantum chaos is still far from being a well understood concept
\cite{FM92}.

In the Heisenberg picture
the classical limit is attained by treating the operator
equations as equations of real numbers, i.e. by neglecting the
non-commutivity of $x$ and $p$ which is of the order of $\hbar$.
The classical limit is equivalent to the limit $\hbar \to 0$.
The advantage of working in this picture is that the classical
limit is directly reached by the above simple prescription. The
disadvantage is that it is difficult to estimate the accuracy of
the classical approach.

In the Schr\"odinger picture, it is useful to consider
wave functions which have minimum spread in both momentum
and space representation. When the system is highly excited,
i.e. the accessible phase space is much larger than the volume of
the wave packet, it describes a state for which
both $x$ and $p$ have ``sharp'' values.
The classical equations of $x$ and $p$ are obtained by calculating
time derivatives of expectation values of $x$ and $p$ under such a
wave function evolving according to the Schr\"odinger equation and
neglecting the corrections due to the finite width of the wave packet. 
These corrections can be shown to be of the order of $\hbar$ and
hence can be
neglected for a highly excited system. The problem in
this representation is that as the system evolves
with time, the minimum uncertainty may not be maintained due
to the influence of interactions. 
The width of the wave packet may start to increase
and the classical equations may become invalid. 
This problem is especially serious for classically chaotic systems.
An exact quantum calculation has been done for the ``Arnold Cat''
system which is chaotic in the classical limit. It was shown
that an initially sharp Gaussian wave packet quickly dissolves into
a diffuse state which is anything but a Gaussian \cite{FMR91}.
Hence we must be especially cautious when we talk about a
localized wave packet in a chaotic quantum system.

Another way to compare classical and quantum mechanics is to
work with the Wigner function $W(p,x)$ which is defined as
\be
W(p,x) = \int^{+\infty}_{-\infty}
dy e^{-ipy/\hbar} \langle x+\frac{1}{2}y|\rho|x-\frac{1}{2}y\rangle.
\ee
 From the Liouville equation we can derive the time evolution of $W(p,x)$,
\be
\frac{\partial W}{\partial t} = ({\cal L}_{c} + {\cal L}_{q})W
\ee
where ${\cal L}_{c}$ is the classical Liouville operator
which describes the classical dynamics in 
phase space language,
\be
{\cal L}_{c} =(\partial_{p} H) \partial_{x} - (\partial_{x} H) \partial_{p},
\ee
while ${\cal L}_{q}$ comprises the quantum corrections

\be
{\cal L}_{q}=\frac{\hbar^{2}}{24}(\partial_{x}^{3}V)\partial_{p}^{3}-
        \frac{\hbar^{4}}{1920}(\partial_{x}^{5}V)\partial_{p}^{5}+...
\ee
which are of successively higher order in $\hbar$.
In principle, in this representation
the quantum corrections to the classical calculation can be computed
order by order in $\hbar$. The rapid dissolution of localized wave
packets for chaotic systems has also been observed in the Wigner
representation \cite{Wi32}, and some attempts have been made to include
quantum corrections by means of a stochastic process \cite{JR87}.
The Wigner function approach has been invoked to derive transport equations
for gauge theories in the mean-field approximation \cite{EH90}, but 
to our knowledge it has not 
yet been applied to gauge theories on a lattice.

\subsection{ Classical Limit of a Gauge Theory}

Now let us turn our attention to nonabelian gauge fields,
where we are dealing with a system of infinitely many degrees of freedom.
What does it mean in this case that the system is highly excited?
Certainly not every degree of freedom can be highly excited, because this
would require an infinite amount of energy.
To this end it is better to look at the
system in Fock space. The system is excited by creating
particles with different energy $E$ and momentum $\hbar k$.
Now suppose the gauge field is excited to a temperature $T$.
The Bose distribution function implies
that the long wavelength particles are more copiously excited.
For $E \ll T$ the Bose distribution $({\rm e}^{E/T}-1)^{-1}$
merges into classical distribution function $T/E$, whereas
the ``hard'' particles of short wavelength are rarely excited. So,
generally, the long wavelength modes can be treated classically
and the short wavelength modes retain their quantum statistics.
With increasing $T$, or decreasing $\hbar$,
more and more modes approach the classical limit. 
If, at certain temperature,
a physical quantity is only related to long wavelength modes, then
we expect the classical calculation to be adequate and the quantum
corrections to be small.

It is here where the lattice regularization of the gauge field plays
an essential role. While being originally advocated \cite{Wi74}
as gauge-invariant cut-off of the ultraviolet divergencies of the
quantum field theory, it can assume a new role in the definition
of the classical high-temperature limit of the quantum field theory.
In the lattice formulation all modes with wavelength
shorter than the lattice spacing $a$ are eliminated,
and we are left only with the infrared modes.
Hence the lattice regulated gauge theory generally goes to the classical
limit at high temperature $T\gg \hbar/a$ 
or vanishing $\hbar$. The validity of the
classical calculation may depend on the nature of the
quantity we are interested in. For example we will show later that the
damping rate for a gluon at rest, as well as the rate of thermalization,
are essentially classical quantities, because they are
induced by the interaction of long wavelength modes.
On the other hand, the  screening length of static electric gauge fields
is controlled by short wavelength modes , and hence is at
best a semi-classical quantity.

The second problem is how to derive the classical field equation from the
quantum field equation. Does the classical equation make any sense?
This question is related to the confinement problem. We can see this
as follows. In the Heisenberg picture we can start with
the operator equation
\be
D_{\mu}F^{\mu \nu}=0,
\ee
where $D$ is the covariant derivative and $F$ is the field operator.
The classical equation is obtained by
treating $F$ and $A$ as numbers rather than operators.
We call $F(x)$ and $A(x)$ classical field configurations.
Suppose now we want to work in the
Schr\"odinger picture. In an unbroken nonabelian gauge theory,
for example in QCD, we observe color confinement so that
only color singlet states are physically realizable. If we calculate
the expectation value of a (color-octet) electric field component 
$E^{a}(x)$ in a singlet
state, the result vanishes according to the Wigner-Eckart theorem.
This means that a classical configuration
does not correspond trivially to a physical state with minimum
uncertainty. The reason is that
a classical gauge configuration is only  gauge covariant while
a singlet state is gauge invariant. But this does not mean that there is
no relation between them. We can start from a classical configuration and
gauge rotate it to obtain other configurations. We then superimpose
these to form a singlet state which corresponds to a physcial state.
This procedure corresponds to the Peierls-Yoccoz projection method
\cite{PY57} for wavefunctions with good symmetry properties.
We can reverse the process and decompose a highly excited color
singlet state into some ``nearly classical'' configurations. By studying these
configurations classically, we can gain insight into the evolution of the
original physical state.

\subsection{Reasons for Investigating the Classical Limit}

The most commonly used approximation
method in quantum field theory is perturbation theory.
But we know that in nonabelian gauge theory there exist some fundamental
non-perturbative effects, such as color confinement
and topological quantum numbers. These effects are
beyond the reach of perturbation theory;
non-perturbative methods are required to understand them. 
The investigation of gauge fields in the classical limit provides
one such approach. Studies of the classical field equations have led
to some very interesting non-perturbative results, such as 
monopole solutions and instantons \cite{HP74,BPST75}. These
classical solutions prove to be vital to the understanding
of the corresponding quantum physics. On the other hand,
these solutions are not general integrals of the classical non-linear
gauge field equations, but exploit special symmetries of the
nonabelian gauge theory. 
In any case, the known classical
solutions of the Yang-Mills equations are by no means exhaustive,
and the equations have been shown to be nonintegrable in general
\cite{MST81}.
Thus we do not have a complete understanding even for the
classical field equations.
It is our hope to learn something about the general
behavior of nonabelian gauge fields by numerically integrating 
the classical equations of motion.

Secondly, under some extreme conditions, e.g. at high temperature, the
quantum field reaches its classical limit at least for some observables.
Quantities that are calculable in the classical limit can be identified
in the high-temperature expansion of perturbation theory as those
that exhibit a leading term proportional to $g^2T$. Namely, if
$g$ is the coupling constant of the classical gauge theory, related
to the standard dimensionless gauge coupling constant $\alpha
=g^2\hbar/4\pi$, the term $g^2T$ has the dimension of an
inverse length or time and survives in the limit $\hbar
\rightarrow 0$. Quantities with this leading behavior are, e.g., the gluon 
damping rate \cite{BP90} and perhaps the inverse screening length of
static magnetic gauge fields \cite{BLS81,BM93}. If we are interested in 
these quantities, the classical approach can provide us with a practical
method of calculating these physical observables.

\subsection{Chaotic Dynamics of Yang-Mills Fields}

The first evidence that Yang-Mills fields exhibit chaotic
dynamics was found a decade ago by Matinyan, Savvidy, and others
\cite{MST81,CS81,NS82,Fr83},
when they studied the dynamics of spatially constant potentials
in the SU(2) gauge theory.  In contrast to electrodynamics, such
potentials are not always gauge equivalent to the trivial vacuum, due
to the non-commutative nature of gauge transformations in nonabelian
gauge theories.  The original motivation for these studies was the desire
to show that the Yang-Mills equations form a non-integrable dynamical system;
it obviously suffices to prove this assertion in a limiting case.
However one can also regard constant potentials as the relevant degrees
of freedom surviving in the infrared limit.  Indeed, L\"uscher has shown
that the Hamiltonian (1) appears as lowest order term in the effective
action for the Yang-Mills theory on a three-dimensional torus, i.e. in a
cubic box with periodic boundary conditions \cite{Lu83}.

Choosing the temporal gauge $A_0^a=0\; (a=1,2,3)$, and assuming that
the vector potentials {\bf A}$^a(t)$ are functions of time only,
the dynamics is governed by the Hamiltonian
\be
H_{\hbox{\fiverm YM}} = \sum_a \textstyle{{1\over 2}}
(\dot {\hbox{\bf A}}^a)^2 + {\textstyle{1\over 4}} g^2
\displaystyle{\sum_{a,b}}
(\hbox{\bf A}^a \times \hbox{\bf A}^b)^2, \label{eq1}
\ee
where $g$ is the gauge coupling constant.  It is not hard to show that
the system allows for seven integrals of motion,
corresponding to energy, angular momentum, and color charge
conservation.  In fact, the Hamiltonian can be reduced to the form
\be
H_{\hbox{\fiverm YM}} \simeq \textstyle{{1\over 2}} (\dot x^2 +
\dot y^2 + \dot z^2) + \textstyle{{1\over 2}} g^2(x^2y^2 + y^2z^2 +
z^2x^2) +\;\ldots \label{eq2}
\ee
where the dots indicate terms describing quasi-rotational degrees of
freedom.  The nontrivial dynamical aspects are all contained in the three
variables $x(t), \; y(t),\;z(t)$.
Note that the coupling constant $g$ can be eliminated by rescaling the
time coordinate.  It is therefore useful to introduce an additional
term into the Hamiltonian which breaks this scale invariance, e.g.,
a harmonic potential:
\be
H_{\hbox{\fiverm YMH}} = H_{\hbox{\fiverm YM}} +
\textstyle{{1\over 4}} g^2v^2 (x^2+y^2+z^2). \label{eq3}
\ee
The dynamical properties of this Hamiltonian are controlled by the
dimensionless parameter
\be
r= \textstyle{{1\over 4}}(gv)^4\big/ g^2E, \label{eq4}
\ee
where $E$ is the energy density.
Numerical studies of Poincar\'e surfaces of section of trajectories
revealed \cite{MST81} that the motion governed by the Hamiltonian 
(\ref{eq3}) is regular
for values $r \gg 1$, becoming partially chaotic as $r$ falls below 1, and
strongly chaotic for $ r\to 0$, as shown in Figure 1.  The mechanism
leading to divergence of nearby trajectories is similar to a classical
billiard:  the equipotential boundary of classical motion for a given
energy $E$ has negative curvature.  A singular-point analysis
of the system (\ref{eq3}) has been performed by Steeb et al. \cite{SLLM86}.
It is not known whether the Hamiltonian (\ref{eq2}) describes a true K-system,
this conjecture has been refuted for the two-dimensional
analogue \cite{DR90}

Space-dependent solutions of the Yang-Mills equation have been studied
in the case of spherical symmetry for solutions of the form
\cite{MPS88,JS89,KO90}
\be
A_i^a = -\varepsilon_{aik} {x_k\over r^2}  (1+\phi(r,t)), \label{eq5}
\ee
which include the so-called Wu-Yang monopole ($\phi\equiv 0$).  The radial
function $\phi(r,t)$ satisfies the nonlinear wave equation
\be
(\partial_t^2 - \partial_r^2)\phi = {1\over r^2}\phi(1-\phi^2), \label{eq6}
\ee
which has been shown to exhibit the rapid energy sharing between Fourier
components which is characteristic of chaotic systems \cite{MPS88,KO90}.
Equation (\ref{eq6}) has also been shown to be non-integrable by the method
of Painlev\'e analysis \cite{JS89}.

Recently, solutions of the Yang-Mills equations in two spatial dimensions
have been studied numerically, subject to the assumption that the
potentials $A_i^a$ depend only on one spatial coordinate and on time
\cite{We92}.  Again mode-sharing of the energy was observed, and 
there are indications that the spatial potential
functions evolve into a fractal pattern.
\bigskip

\section{Chaos in SU(2) Lattice Gauge Theory}

\subsection{General Considerations}

Interesting as these results are, they leave two important questions
unanswered:  What are the dynamical properties of the full (3+1)-dimensional
classical Yang-Mills field?  What is the physical significance of chaotic
dynamics of the classical field theory?  The goal of our investigation
was to provide at least partial answers to these questions.  Our approach
\cite{MT92} deviates in two important aspects from earlier studies:

(a) Since the spatial coordinates have to be discretized for numerical
purposes, it is convenient to formulate the SU(2)-gauge theory in terms
of matrix-valued link variables on a $N^3$ cubic lattice 
\cite{Wi74,Cr83}: 
\be
U_{x,i} = \exp \left( -\textstyle{{1\over 2}} iga A_i^a (x)\tau^a
\right). \label{eq7}
\ee
Here $\tau^a$ are the Pauli matrices, $a$ is the elementary lattice
spacing, and $(x,i)$ denotes the link from the lattice site
$x$ to the nearest neighbor in direction $i$,  $x+i$.  The link
variables $U_{x,i}$ are explicitly gauge covariant, as opposed to the Yang-Mills
potentials $A_i^a(x)$.  Since the $U_{x,i}$ take values on the gauge
group SU(2) rather than the group algebra, the magnetic field
strength is bounded for a given lattice spacing.

(b) Instead of studying the gauge field dynamics in the vicinity of
arbitrarily selected configurations, we have investigated the dynamical
behavior of {\it random} field configurations, corresponding to gauge
fields selected from a microcanonical or canonical ensemble.
This has the advantage that our
field configurations are controlled by a single parameter, the
temperature $T$ or the average energy density $\varepsilon$, which can
be varied systematically.  This approach also allows for the identification
of some quantities calculated for the classical Yang-Mills theory with
those obtained in the high-temperature limit of the quantum field
theory.

\subsection{Lattice SU(2) Theory in Hamiltonian Formalism}

Our study is based on the Hamiltonian formulation of lattice
SU(2)-gauge theory \cite{KS75,CRUK85}, governed by the Hamiltonian
\be
H = {a\over g^2}
\displaystyle{\sum_{x,i}} \tr({\dot U}^{\dagger}_{x,i}{\dot U}_{x,i}) +
{4\over g^2a} \displaystyle{\sum_{x,ij}}
[1-\textstyle{{1\over 2}} \tr U_{x,ij}] ,
\label{eq8}
\ee
where a dot denotes the time derivative. Here
the electric and magnetic fields have been
expressed in terms of the SU(2) link variables $U_{x,i}(t)$
and the so-called plaquette operator $U_{x, ij}$ which is the product
of all four link variables on an elementary plaquette with corners
$(x, x+i,x+i+j,x+j)$:
\be
U_{x,ij} = U_{x,i} U_{x+i,j} U_{x+i+j,-i} U_{x+j,-j}
\ee
with $U_{x,-i} = U^{\dagger}_{x-i,i}$.
The links are directed and hence the plaquettes are oriented.
In the continuum limit the plaquette variable $U_{x,ij}$ is related
to the local magnetic field $B_{x,k}^a$
\be
U_{x,ij} = \exp(-i\oh g a^2 \epsilon_{ijk} B^a_{x,k} \tau^a), \label{eqB}
\ee
while the electric field on the lattice is given by
\be
E^a_{x,i} = - \frac{ia}{g^2} \tr (\tau^a \dot U_{x,i} U^{\dagger}_{x,i}).
\label{eqE}
\ee
In the classical limit, this Hamiltonian is scale invariant.
 To see
this explicitly, we scale the time variable, $s=t/a$, obtaining
\be
g^2 Ha = \sum_{x,i} \hbox{tr} \left( {\delta U_{x,i} \over
\delta s}\; {\delta U_{x,i}^{\dagger}\over \delta s}\right)
+ 4\sum_{x,ij} [1-\textstyle{{1\over 2}} \tr U_{x,ij}], \label{eq13}
\ee
where the right-hand side is parameter free. So the only parameter
in the system is total energy or temperature.

The classical equations of motion are derived from this Hamiltonian,
making use of the lattice representation of the electric field
components (\ref{eqE}). It is useful to write the SU(2) matrices
in the form
\be
U = u_0 - i \tau_a u_a = \left (
    \begin{array}{rr}
	u_0-i u_3,& u_2 - i u_1 \cr
	-u_2-i u_1,& u_0 + i u_3 \cr
    \end{array} \right )
\ee
where the $u_i$ are four real numbers, which can be thought of as components
of a quaternion. The unit determinant implies that
\be
\det U = u^2_0 + u^2_1 + u^2_2 + u^2_3 = 1.
\label{length}
\ee
One easily verifies that the components of the quaternion satisfy the
following differential equations of motion:
\be
(\dot u_0)_{x,i} = \frac{g^2}{2a} E^a_{x,i} u^a_{x,i}
\label {U0dot}
\ee
and
\be
(\dot u_a)_{x,i} = \frac{g^2}{2a} \left [
E^a_{x,i} (u_0)_{x,i} + \epsilon^{abc} E^b_{x,i} u^c_{x,i} \right ].
\label{Udot}
\ee
These conserve the quaternion length, $||U||=\det U$, because they satisfy
\be
\dot u_0 u_0 + \dot u_a u_a = 0.
\ee
Although it would be sufficient to update only the three
fields $u_a$ during the time evolution because of the
unit length constraint, it is computationally more efficient to also
update the fourth component $u_0$, as well using
eq. (\ref{U0dot}) rather than calculating $u_0$ by taking the
square root of $(1-u_au_a)$. The three electric fields $E^a_{x,i}$
on each link are updated according to
\be
\dot E^a_{x,i} = \frac{i}{a g^2} \sum_{j}
\tr [\oh \tau^a (U_{x,ij} - U^{\dagger}_{x,ij})]
\label{Edot}
\ee
where the sum runs over all four plaquettes that are
attached to the link $(x,i)$. Here $\tr (\oh \tau^a Q)$,
for any quaternion $Q$, just reads off the component
$q^a$, and therefore does not require additional computational effort.

We note here that the time evolution for the electric field
conserves Gauss' law
\be
D^{ab}_i E^b_{x,i} = 0,
\label{Gauss}
\ee
which is an expression of charge conservation.
Starting with any configuration $(U_{x,i}, E_{x,i})$
that satisfies Gauss' law, the time evolution will not violate it.
Precisely because of the chaotic nature of Yang-Mills dynamics it is,
in general, impossible to integrate this set of
equations analytically. We have to rely on numerical methods, which
will be discussed in the following subsection.

\subsection{Integrating the Equations of Motion}

\subsubsection{Lattice Geometry and Boundary Conditions}

In all our simulations we discretize the gauge
fields on a simple cubic $N^3$ lattice with periodic boundary
conditions. To calculate the oriented
plaquette products more efficiently we use a linked-list approach.
All the sites are numbered consecutively.
The link numbers are in $x$, $y$ and $z$ order, and are arranged
according to the site of origin. The plaquette numbers are
chosen so that they are orthogonal to the links in $x$, $y$ and $z$
directions, respectively. We also build two additional
arrays of numbers at the start of the simulation. One contains 
the link number of all the links that form a given plaquette,
ordered in the way they appear in the directed
plaquette product. The other list contains the number
of plaquettes that contain a given link, ordered
according to the position of the link in the plaquette.
The periodic boundary conditions are reflected automatically
in the above two link lists. 

\subsubsection{SU(2) Representation}

There are several ways to represents a SU(2) matrix:
by a complex $2 \times 2$ matrix, by a real quaternion, or in polar
coordinates as a point on the 4-dimensional unit sphere
which can be specified by 3 angles.
The last representation has the lowest storage requirements,
however, it involves time-consuming trigonometric conversions.

In our simulations we chose the quaternion representation
because it takes the least number of floating point operations
to multiply two group elements. In the complex
matrix representation one needs 32 multiplications and 24
additions compared to 16 floating point multiplications
and 12 additions when taking the product of two quaternions.
Moreover, the equations of motion can be written
in a simpler form in quaternion representation.

\subsubsection{Numerical Integration}

The numerical task consists in propagating the gauge fields in time,
by integrating the equations of motion.
This can be achieved by a variety of numerical methods.
Here we use the Runge-Kutta method with fourth-order accuracy,
which is easy to implement, allows for adjustable time-step
control, and is quite stable.

The most CPU intense part of the simulation is the computation
of the oriented product of the fields over the complement lattice.
This complement lattice is defined by all oriented links contained
in the elementary plaquette attached to a given link.
The overall performance of the code depends on several issues
like field representation, fast access to the data on the neighboring
links and plaquettes, etc. 

\subsubsection{Code Verification and Accuracy}

Several integrals of motion can be used to
verify the simulation code and to test the accuracy of
the time integration.
Since we are studying the Hamiltonian evolution of
the gauge fields, the total energy of the system
remains constant. With a typical value for the time step
of the  order of 0.01 dimensionless units,
the total energy is conserved to better than 8 significant digits.

Another conserved quantity is the length of each quaternion
link variable for SU(2).
For an interval less than one dimensionless time unit the
conservation is better than 12 significant figures. However,
due mainly to accumulation of cut-off and rounding
errors the precision is deteriorating progressively.
Therefore we choose to rescale the
gauge fields after every time integration
to maintain fixed length of each dynamical variable, since the subsequent
integrations are very sensitive to the quaternion length preservation.
This method permits to integrate the equations of motion with
a larger time step.
Later we will also study SU(3) gauge theory,
where the corresponding criterion is that the determinant of the
unitary matrix on each link shall be unit. In this respect,
the SU(3) evolution is more stable. The deviation of the
determinant from one is of order $10^{-8}$ even if the system evolves
over a total time interval of $T=20$ or 30.

The validity of Gauss' law, eq. (\ref{Gauss}),
is also an indication for accurate integration, which is extremely
sensitive to all kinds of program errors and thus provides a
valuable probe for code verification. In our calculations
the color charge was always conserved to better than five
significant digits.

\subsubsection{Performance}

The most time consuming part of the code is the
calculation of oriented plaquette products.
However, this part of the code is fully vectorizable
and runs at about 160 Mflops for SU(2) and
100 Mflops for SU(3) on a single Cray-Y-MP processor.
Typically a single time-step integration of the set
of equations of motion for $N=10$ takes about 30 ms CPU time
for SU(2) and 220 ms for SU(3)
on the same processor.

\subsection{Divergence of Trajectories}

In order to look for chaotic motion we will consider the evolution of
infinitesimally separated gauge field configurations.  If we find
exponentially diverging trajectories, we can identify the positive
Lyapunov exponents and obtain the value of the Kolmogorov-Sinai
entropy, i.e. the entropy growth rate,
of the gauge field. 
To observe the exponential
divergence of two trajectories $U_{x,i}(t)$ and $U'_{x,i}(t)$
in the space of gauge field configurations
we introduce the following gauge-invariant distance:
\be
D[U_{x,i},U'_{x,i}]
= {1 \over 2 N_{\rm p}} \sum_{x,ij} | \tr U_{x,ij} - \tr U'_{x,ij} | ,
\label{eq10}
\ee
where $N_{\rm p}=3N^3$ is the total number of elementary plaquettes.
In the continuum limit
\be
D[U_{x,i},U'_{x,i}] \buildrel {a\to 0}\over\longrightarrow
D[{A'}_i^{a},A_i^a] 
\propto {1\over 2 V} \int d^3x \vert B'(x)^2 - B(x)^2\vert,
\label{eq11}
\ee
i.e. $D$ measures the average 
absolute local difference in the magnetic energy of
two different gauge fields. We note a
peculiar property of this distance measure, which is a natural
consequence of the topology of the compact $9N^3$-dimensional space
of magnetic gauge field configurations on the lattice: 
For $N \gg 1$ almost all pairs of configurations have the same distance $D$.
This is illustrated in Figure 2, where the distribution of distances
of randomly chosen configurations from a fixed field configuration
is shown for lattices of different size. For large lattices the
distribution approaches a narrow Gaussian with a
width of order $N^{3/2}$.
This property does not limit the usefulness of the metric (\ref{eq10})
as measure of the divergence of infinitesimally separate field
configurations, but it causes the saturation of $D$ at large times
observed in the calculations (see figures below).

Figure 3 shows the evolution of $D(t)$ for initially neighboring
gauge field configurations on a $20^3$ lattice. We choose the
reference configuration by randomly selecting link variables in
such a way that the average energy per plaquette takes on the
desired value \cite{MT92}. This procedure is controlled by a parameter
$\delta$ which varies between 0 and 1. The energy per plaquette grows
like $\delta^2$ for small $\delta$ and saturates in the limit $\delta
\to 1$.  We then construct a neighboring configuration by perturbing
each link element infinitesimally (see \cite{MT92} for details).
For values of $\delta$ of order unity the distance $D(t)$ starts
to grow exponentially as $D(t) = D_0 \exp(h t)$ almost immediately
(see Figure 3a). The growth rate $h$ decreases with $\delta$, and
for $\delta \ll 1$ one observes an extended period during which
the distance $D(t)$ between two adjacent field configurations
performs more or less regular oscillations before exponential growth 
finally sets in (see Figures 3b, 4). For very small values of $\delta$,
corresponding to very low energy density of the gauge field
configurations, the exponential growth pattern of $D(t)$ is
modulated by low-frequency oscillations, which we attribute to the
growing influence of non-leading Lyapunov exponents.

The initial latency period before the onset of exponential growth of
$D$ can be estimated as follows:  We can write $U'_{x,i}(t) = U_{x,i}(t) +
\delta U_{x,i}(t)$ where  $\delta U_{x,i}$ approximately satisfies a 
system of linear
differential equations with $18N^3$ eigenmodes and eigenfrequencies.  
On average every eigenmode will be excited with equal probability by our
random choice of $\delta U_{x,i}(0)$.  In order for the maximally
unstable mode to outgrow the combined weight of all other modes we
therefore have to wait a certain time $t_0$, statistically given by
$\exp (ht_0) \approx (18N^3)^{1\over 2}$, i.e.
\be
t_0 \approx \ln (18N^3)/2h. \label{eq12}
\ee
This agrees roughly with the observations, in particular, it
explains why the onset of exponential growth is delayed for small values
of the slope parameter $h$ (see Figure 4).

\subsection{Maximal Lyapunov Exponents}

It is natural to identify the growth rate $h$ with the maximal Lyapunov
exponent $\lambda_0$ of the lattice gauge theory. This is confirmed
by a careful analysis of the Lyapunov spectrum (see sections 3.6 and 3.7 
below), and we will assume the equality $h=\lambda_0$ in the following.
Extensive studies have shown that $\lambda_0$ is a universal
function of the average energy per plaquette $E$, as shown in Figure 5.
For values $\delta > 0.15$, the numerical determination of $\lambda_0(E)$
from $D(t)$ is quite reliable, and the statistical and systematic
errors are small. Figure 5 demonstrates that $\lambda_0(E)$ is growing
approximately linearly with energy.  
Using the property of scale invariance of the Hamiltonian, discussed 
previously in eq. (\ref{eq13}), one finds that the dimensionless product
$\lambda_0 a$ can only be a function of the combination $g^2Ea$. 
Our numerical results show that this function is approximately linear:
\be
\lambda_0 a \approx \textstyle{{1 \over 6}} g^2 E a . 
\label{eq14}
\ee
This scaling property has been verified numerically over a wide
range of values for $g$ and $a$ (see Figure 5). We note that, according 
to (\ref{eq14}), $\lambda_0$ is independent of the lattice spacing $a$ 
in the classical limit, where $g$ does not depend on $a$.

We have also studied the dependence of $\lambda_0(E)$ on the size of the
lattice at fixed lattice spacing $a$.  Figure 7 shows the evolution
of the distance between adjacent field configurations for lattices
of size ranging from $N=6$ to $N=28$.  The rapid convergence is
obvious, the curves for $N=28$ hardly deviate from those for $N=6$,
except for a decrease of fluctuations which is not visible in the
figure because the curves almost coincide.
We have not observed within statistical errors a systematic dependence
of $\lambda_0(E)$ on $N$, for $N \ge 6$ and $\delta > 0.15$. The results
obtained for $N=6$ are shown in Figure 5 as solid squares. 
For smaller values of $\delta$ we found that 
the exponential growth rate gradually decreases with growing $N$,
which may explain why the lowest two points in Figure 5, obtained
for $N=20$, still lie above the straight line.

\subsection{Rescaling Method for Lyapunov Analysis}

The previous method for obtaining the largest Lyapunov exponent 
of a Hamiltonian system is straightforward, but it has two drawbacks. First, 
the exponential divergence of trajectories shows fluctuations, resulting 
in an uncertainty in the determination of the exponential divergence rate.
The second drawback of the method is that only the largest Lyapunov
exponent can be obtained in this way, but not the complete Lyapunov spectrum.
We now discuss a method which can be used to determine Lyapunov 
exponents more precisely and yields the whole
spectrum of Lyapunov exponents.

This technique, which we call rescaling method here, is widely used in
studying chaotic dynamical systems \cite{BFS79}.
Suppose we want to calculate the two largest
Lyapunov exponents of the system. We can randomly choose three initial
points in phase space, to which we refer as $z_0(0),z_1(0),z_2(0)$ for
convenience, with the condition that they are
close to each other according to some appropriate distance measure.
If the system is chaotic, or if the initial points are chosen inside
the chaotic part of the phase space, the distances between
the three trajectories $z_{i}(t)$ evolved
from the three initial points will diverge exponentially. Let us denote
the separation vectors between the trajectories by
\be
d_i(t) = z_i(t) - z_0(t),\qquad (i=1,2)
\ee
and the absolute distance between $z_i$ and $z_0$ by $D_i=|d_i|$.
Since the available phase space volume is limited by the total available 
energy, $D_i(t)$ will saturate after a certain time.
To avoid saturation, the following rescaling method is used.
The fixed reference distance $D_0 = D_i(0)$ is chosen in the beginning.
The whole procedure consists of two steps. In the
first step, the trajectories $z_{i}(t)$ evolve according to the equations
of motion for a period of time $t_{0}$. Then in the second step we
rescale the separation vectors $d_{1}$ and $d_{2}$ as follows.
We hold the point $z_{0}(t_0)$ fixed, but scale the
distance $D_1$ back to $D_{0}$ by setting:
\be
z'_1(t_0) = z_0(t_0) + {D_0 \over D_1(t_0)} d_1(t_0).
\ee
The scaling factor is denoted by
\be 
s_{1}^{k} = D_1(t_0)/D_0, 
\ee
where $k$ refers to the $k$th rescaling. 
For $z_{2}(t_0)$, we first orthogonalize $d_{2}(t_0)$ against $d_{1}(t_0)$, 
then scale the orthogonalized vector $d'_{2}$ to the reference length 
$D_{0}$.  The new scaling factor is denoted by $s_{2}^{k}$. 
This procedure is iterated $n$ times until the Lyapunov exponents,
\be
\lambda_{i}=\lim_{n\rightarrow \infty}
        \sum_{k=1}^{n}\frac{\ln s_{i}^{k}}{t_{0} },
\ee
converge.
Here $\lambda_{1}$ is the largest Lyapunov exponent and $\lambda_{2}$
is the second largest one. By adding more and more trajectories, 
in principle, this method can be used to
obtain the whole Lyapunov spectrum with arbitrary accuracy.
The time needed to obtain a given Lyapunov exponent
depends on how fast the procedure converges.

At first glance, the applicability of this method to lattice gauge
theory is questionable. The reason is that while the meaning of
rescaling and
orthogonalization is quite obvious in a Euclidean phase space, it is
less clear in the case of a gauge field,
where the phase space is curved as well as constrained by Gauss' law.
The problem of curvature is relatively simple. One possible
approach is to transform the link variable $U_{x,i}$
back to the vector potential $A_i^a$, and to work in the phase space
formed by electric fields and vector potentials,
for which the geometry is Euclidean.
In the case of SU(2), we have yet a simpler method. Each SU(2)
group element is represented by a normalized quaternion.
When performing orthogonalization and rescaling, we can simply
treat all components of the quaternion as independent cartesian
coordinates, because locally the metric of the curved space is Euclidean.

Surprisingly, Gauss' law also does not pose a serious threat, for
the following reason. When we rescale the separation vectors $d_i$, 
we indeed violate
Gauss' law. But if the distance $D_i$ is small, then the violation
of Gauss' law is of second order in $D_i$. If we limit ourselves
to sufficiently small $D_i(t_0)$, then the violation of Gauss' law in each 
rescaling step is negligible.
We next observe that the evolution of the system respects Gauss' law, 
so the violation does not increase with time. On the other hand, the
next rescaling decreases the previous violation of Gauss' law
by a (large) scale factor, so the violations do not accumulate. The
same argument applies to the small changes in the choice of gauge 
induced by the linear rescaling procedure.

We have used this method to study the SU(2) theory and have measured
the largest two Lyapunov exponents.
We indeed find that the violation of Gauss' law remains 
of the order of $10^{-6}$. The result of a typical run for the 
Lyapunov exponents is shown in Figure 8 
for a configuration with scaled energy $g^{2}Ea=4.06$.
The solid line corresponds to the largest exponent and the dotted line to 
the second largest one. They converge at $t \approx 100$. Note the time scale
of saturation of the distance $D(t)$ in the case without rescaling is 
about 30 at the same energy. The result obtained with our new, improved
method, $\lambda_{0}=0.667$, is very close to, 
but slightly lower than, our previous result $4.06/6=0.677$,
where we did not use the rescaling technique. We note that the results
for the Lyapunov exponents generally converge from above, i.e. the
Lyapunov exponents are overestimated when the trajectories are not
followed for sufficiently long time.  We also observe 
that $\lambda_{1}$ is almost identical to $\lambda_{0}$. 
The reason is that, as we are going to show next, there exists a whole
Lyapunov spectrum which forms a continuous curve in the thermodynamic limit. 

\subsection{Lyapunov Spectrum}

The above method, in principle as well as in practice,
can be extended to obtain the whole Lyapunov spectrum for small lattices.
For each additional Lyapunov exponent, we need to integrate one more 
trajectory so as to form one additional linearly independent vector $d_i$.
If we want to calculate $\nu_L$ Lyapunov exponents, we need to run
$\nu_L+1$ configurations simultaneously, from which we can form $\nu_L$ 
separation vectors and calculate the $\nu_L$ largest Lyapunov exponents.
Practically, the computational resources put limits on the possible
size of the lattice for which we can obtain the whole Lyapunov spectrum.
We have performed studies with $1^3$, $2^3$ and $3^3$ lattices.
Fortunately, we found that the spectrum starts to scale as early as size 
$3^3$, which permits us to extrapolate the result to the thermodynamic 
limit.

We have calculated the complete spectrum for $1^3$ and $2^3$ lattices.
The results for the latter are shown in Figure 9. 
On a $N^3$ lattice the dimension of phase space is
$9N^3\times 2$, 
because there are three vector and three color directions at each site
for magnetic and electric fields. This amounts to a total number 
of 144 Lyapunov exponents for $N=2$. 
We can see that the spectrum is divided into three equal parts.
The first one third of Lyapunov exponents
are positive, while the second one third are all zero, and the last 
one third of exponents are the negative of the first one third.
The vanishing Lyapunov exponents account just for all the 
degrees of freedom associated with static gauge 
transformations and with conservation of Gauss' law, each equal in number
to three times the number of lattice sites. Thus
our results confirm the general properties of Lyapunov spectra\cite{BFS79}.
The results for a $1^3$ lattice are basically the same, but the Lyapunov
spectrum consists only of 18 numbers. The value of the largest Lyapunov 
exponent is consistent with the result obtained earlier for the model
of spatially constant Yang-Mills potentials (see section 2.4).

In Figure 10 we show the scaling of the Lyapunov spectrum,
where we compare the results from a $2^3$ lattice and a $3^3$ lattice.
To save computation time, we have calculated only the positive Lyapunov 
exponents for the $3^3$ lattice. In both cases, the
initial configurations are chosen similarly. For presentation 
the Lyapunov exponents were scaled with respect to the largest 
exponent $\lambda_0$ in each case.
The indices labeling the Lyapunov exponents are scaled to the total
number of Lyapunov exponents, i.e. 144 for a $2^3$ lattice and 486
for a $3^3$ lattice. The two lines coincide nicely, exhibiting an early
scaling behavior. 

The scaled Kolmogorov-Sinai entropy, i.e. the sum
over all positive Lyapunov exponents, is:
\be
\frac{\sum_{i}\lambda_{i}}{N^{3}\lambda_{0}} \approx 2,
\ee
where $N^3$ is the size of the lattice.
We point out that at this small lattice size, the spectrum does not
yet scale with energy. But we know from section 3.4 that
this scaling appears at a lattice of size $6^3$,
where we found $\lambda_{0}/g^2E \approx \frac{1}{6}$.
If we conbine both scaling laws, we can estimate the Lyapunov spectrum
in the thermodynamic limit, from which we obtain the Kolmogorov-Sinai
entropy density
\be
{\dot\sigma}_{\rm KS} = {1\over (Na)^3} \sum_i \lambda_i 
\approx \frac{1}{9}g^{2}\varepsilon,
\ee
where $\varepsilon=3E/a^{3}$ denotes the energy density,  
and $E$ is the average energy per elementary plaquette.
Makinmg use of the thermodynamic relation 
$\sigma T = \varepsilon + P = {4\over 3}\varepsilon$,
we find that the characteristic entropy growth rate for SU(2) is
given by
\be
{{\dot\sigma}_{\rm KS}\over\sigma} 
\approx {1\over 9} g^2 {\varepsilon\over\sigma}
= {1\over 12} g^2 T .
\ee

The above method is quite successful for SU(2) gauge theory, but
it is not obvious how to apply it to SU(3) gauge fields. The reason is
that this method relies on the quaternion representation, which is
quite special for the group SU(2). We have also developed
a more general method to obtain the Lyapunov spectrum, which
can be used to study SU(3) gauge theory. Again the basic idea is simple. 
We want to directly consider the motion of vectors in the tangent
space built upon the phase space \{$E_{x,i},U_{x,i}$\}. The tangent space 
of $E_{x,i}$ is simple, the vectors just being given by $\delta E_{x,i}$.
We must be more careful specifying a vector in the
tangent space to $U_{x,i}$. Here, to be consistent with our
definition of the conjugate momenta $E_{x,i}$ as the left group generators,
we define a vector $b_{x,i}$ in the tangent space of $U_{x,i}$ as
$\delta U_{x,i}=i b_{x,i} U_{x,i}$. A vector in the complete tangent 
space is the direct product of $b_{x,i}$ and $\delta E_{x,i}$. 
The linear evolution equations for $b_{x,i}$ and $\delta E_{x,i}$
can be derived and integrated along with
the equations of motion in phase space. Initially choosing 
$\nu_L$ vectors, we can again obtain $\nu_L$ Lyapunov exponents.
The results for SU(2) by this method is shown in Figure 11, in comparison
with the result obtained by the previous method. We observe good agreement
for large positive Lyapunov exponents, but increasing
deviations for the smaller ones.
These deviations can be eliminated by demanding higher
numerical precision for the second method.

\section{Other Lattice Field Theories}

In this section, we want to apply our method to several other interacting
fields.  We calculate their largest Lyapunov exponents and investigate 
how they scale with energy, in order to determine the nature of the 
stochatic behavior of the various field theories.

\subsection{U(1) Gauge Theory}

How characteristic is the observed behavior for the SU(2) gauge
theory? Of course, one would expect a similar chaotic behavior for higher
SU(N) gauge theories, because SU(2) is a subgroup of SU(N).
However, one may expect a completely different behavior in the case
of compact U(1), which has no self-interaction of gauge bosons
in the continuum limit.
The gauge group U(1) is easily obtained as a one-parameter
abelian subgroup of SU(2).  Not unexpectedly, we find quite different
results in this case.  For small values of the parameter $g^2Ea$ there is no
discernable exponential divergence of adjacent field configurations.
As shown in Figure 12, $\lambda_0a$ suddenly begins to grow rapidly for 
values $g^2Ea \approx 2$.  However, the dependence is highly nonlinear,
reminiscent of the behavior of many nonlinear dynamical systems.
This proves that the approximately linear relationship between 
$\lambda_0 a$ and $g^2Ea$ found for the
gauge group SU(2) is far from trivial.  The latter is obviously 
associated with the nonabelian nature of this group.  It is also worthwhile
noting that $\lambda_0\to 0$ for $a\to 0$ in U(1), i.e. the maximal 
Lyapunov exponent $\lambda_0$ for the compact U(1) lattice gauge theory 
vanishes in the continuum limit.

\subsection{SU(3) Gauge Theory}

Since SU(3) contains SU(2) as a subgroup, we will certainly
expect that SU(3) gauge thoery is chaotic.
This expectation is borne out by our study of the dynamics
of SU(3) gauge fields on a lattice, which shows the same type of exponential
instability \cite{Go93}. While the basic procedure here is same as in
SU(2), technically SU(3) is more complicated than SU(2).
But nevertheless, since SU(3) is physically relevant as the gauge theory
of the strong interaction, it is worthwhile to study its
general dynamical properties. 

Whereas a SU(2) group element can be represented
by a quaternion with only one redundant variable, there is
no such simple representation for a general SU(3) group
element. In principle we can use the exponential representation
with 8 angles as independent variables. But as in the case of SU(2) 
this would involve many time consuming trigonometric
manipulations required to carry out a group multiplication.  In our 
simulations we have, therefore, directly used the matrix representation,
i.e. we represent every group element by a unitary $3\times 3$ matrix.
With the cost of 10 redundant variables for each link
the group multipliaction is directly realized by matrix
multiplication which only involves multiplications and additions.

Another difference we shall point out is 
that it was necessary to be more careful in choosing the initial
states here than in the case of SU(2), 
because the topological structure of the group
space of SU(3) is more complicated.  SU(3) is an eight-parameter Lie
group and not all directions in group space are equivalent.  Simply
choosing angles at random in a certain representation of the group parameters
\cite{Br88} can lead to a sampling of gauge fields that differs strongly from
the thermal ensemble.  It therefore proved necessary to choose the initial
state by the heat-bath method \cite{Cr83,Pi81}, which is straightforward but
numerically expensive especially at low temperatures where the rejection
rate is large.  All other considerations, especially the scaling property
(\ref{eq13}), carry over to SU(3).

Our numerical results, obtained by evolving the thermalized
initial condition, are presented in Figure 13. They show again a close to
linear dependence between $\lambda_0$ and $g^2E$, but with a different 
slope than for SU(2).  To good approximation we find:
\be
\lambda_0 a \approx \textstyle{{1\over 10}} g^2 Ea.
\label{eq15}
\ee
We investigated several cases with non-thermalized initial conditions,
specified by restricting some angles to a certain limited range.
These initial conditions lead to exponential divergence rates
lying above the line (\ref{eq15}). We surmise that this reflects the
lack of complete randomization of the field configuration during the
limited time interval until the distance measure saturates.  This suggests 
the importance of the thermalized initial condition, when the rescaling
method is not applied.  We have also tried a different definition
of distance
\begin{equation}
D^{E}[U,U'] =\frac{1}{N_{\rm p}} \sum_{x} \mid \sum_{i}
\tr({\dot U}^{\dagger}_{x,i}{\dot U}_{x,i}) -
\tr({\dot U}'^{\dagger}_{x,i}{\dot U}'_{x,i}) \mid,
\end{equation}
corresponding to the sum of the absolute value of local difference in the
electric energy. The rise of $\ln D^{E}(t)$
is coincident with that exhibited by $\ln D(t)$
except for the initial oscillatory region.
This is within our expectation because the chaoticity of a system is an
intrinsic property and it does not depend on a particular choice
of the distance measure.

\subsection{Massless Scalar Field}

One question concerning the chaoticity of nonabelian gauge fields
is whether the chaoticity is just a consequence of the nonlinearity of 
the field equations, or whether it is related to the particular form
of the nonabelian gauge interaction.
In this respect, we study the classical $\Phi^{4}$ theory
described by the Lagrangian
\be
L = \partial_\mu \Phi^{\dagger} \partial^\mu \Phi
-\mu^2 \Phi^{\dagger} \Phi - g^2 (\Phi^{\dagger} \Phi)^2.
\ee
The lattice formalism for interacting Higgs and
gauge fields was given by Ambj\o rn et al. \cite{AAPS92}.
We here study a special case, i.e. the massless limit of
an iso-doublet complex scalar field, which has four real field components,
without the gauge field. In the massless limit ($\mu=0$),
just like in the case of a gauge theory, the corresponding
classical lattice theory is scale invariant,
in the sense that the self-coupling $g^{2}$ of the scalar field
and the lattice spacing $a$ can be scaled out entirely and
the system has no free parameter except the total energy.

Using the same method as before, we measured Lyapunov
exponents of the massless scalar field on the lattice. In
Figure 14 we show $\lambda_0/ g^{2}E$ as function of $g^{2}Ea$.
Observe that the maximal Lyapunov exponent for the scalar field
is much smaller than that for gauge fields with the same
energy. Second, we find that the ratio $\lambda_0/ g^{2}E$
tends to zero at small $g^2Ea$, suggesting a vanishing Lyapunov exponent
in the continuum limit $a \rightarrow 0$. This result is consistent with 
our understanding of the relation between the maximal Lyapunov
exponent and the damping rate of the long wavelength modes.
The thermal perturbation theory calculation shows,
unlike in the case of nonabelian gauge fields, that
the damping rate of the long wavelength mode of the scalar field vanishes
at the order of $g^{2}$. This result shows that not all nonlinear
classical continuum field theories are chaotic. 

\subsection{Massive SU(2) Vector Fields}

This is the first step toward the understanding of chaoticity
in  the electroweak interaction, where the gauge symmetry
is spontaneously broken by a Higgs field. We want to study how a
mass term affects the chaotic behavior, in order to find a
hint of how the interaction between the Higgs field and
the gauge field may affect the chaoticity of the latter.
In Figure 15, we show the largest Lyapunov exponents at various 
energies for theories with different vector boson mass.  The same 
scaled variables are used as before, but here the mass parameter $m$
cannot be scaled out.
It is related to the mass $M$ of the vector field
in the continuum theory as $M=m\hbar/2a$.  For comparison, 
the results for the massless gauge theory are shown as solid squares,
which are fit by a straight line with a coefficient of ${1\over 6}$.
The hollow squares are for $m=0.2$, and the crosses are for $m=4$.
They more or less lie on a straight line. We observe that the effect
of a mass term is to reduce the chaoticity of the system, or in other
words, the mass term has a stabilizing effect on the trajectories
of the field configurations. This effect is consistent
with studies of the simple system (\ref{eq3}),
where the amount of chaoticity depends on the relative strength
of the nonlinearity and the harmonic potential \cite{MST81}.

\subsection{Spontaneously Broken SU(2) Yang-Mills Theory}

The pure SU(2) and SU(3) Yang-Mills gauge theories
are the primary tools for studying nonperturbative QCD related
phenomena, such as properties of a hot quark-gluon plasma.
The spontaneously broken SU(2) Yang-Mills theory, in which a charged
scalar isodoublet field, the Higgs field, is coupled to the gauge boson,
is a model of the electroweak gauge theory. It
is used to study the high-temperature limit of the standard model
of particle physics, in particular, the phenomenon of baryon number
violation at temperatures around the electroweak phase transition
temperature of $T=200$ GeV.

In the limit of vanishing Weinberg angle,
the Hamiltonian describing this model is given by
\be
H = \int d^3x \left [
\oh E^a_i E^a_i + \frac{1}{4} F^a_{ij} F^a_{ij} +
\dot \Phi^{\dagger} \dot \Phi + (D_i \Phi)^{\dagger} (D_i \Phi) -
\mu^2  \Phi^{\dagger} \Phi + \lambda (\Phi^{\dagger} \Phi)^2,
\right ]
\label{H-Higgs}
\ee
where the dot symbol denotes time derivative,
\be
\Phi = \left (
  \begin{array}{c}
	\phi_1+i \phi_2 \cr
	\phi_3+i \phi_4 \cr
  \end{array} \right )
\ee
is a charged Higgs doublet in the fundamental representation of SU(2),
and the gauge field is described by vector potentials $A_i^a(x)$
in the temporal axial gauge $A_0^a=0$.
In the numerical simulation we again discretize the gauge fields 
and represent them by link variables $U_{x,i}$, which are identified
with real-valued quaternions.
The Higgs doublet is also represented as a quaternion. 
However, its length (determinant) is also a
dynamical degree of freedom, which we denote by $R$:
\be
R^2 = \oh \tr (\Phi^{\dagger} \Phi).
\label{R}
\ee
Following the notation of Amj\o rn et al. \cite{AAPS92},
we introduce a unit length quaternion $V$ for the representation
$\Phi = RV$.
This factorization is a useful way to separate the gauge-invariant
and the gauge-dependent degrees of freedom of the Higgs fields $\Phi$.
Nevertheless, in most formulas we shall use the
matrices $\Phi$, $\Phi^{\dagger}$, for brevity.

The lattice regularized version of the Hamiltonian (\ref{H-Higgs})
is based on the link-variables $U_{x,i}$, defined as in eq. (\ref{eq7})
and the site-variables, $\Phi_x$, or equivalently $R_x V_x$, 
for the rescaled Higgs field matrix
\be
\Phi_x = \frac{a}{\sqrt{\beta_H}} \Phi(x) .
\ee
The Hamiltonian consists of three terms. The first one
is the same as for the pure Yang-Mills theory
\be
H_{\rm W} = {\beta_{\rm G}\over a} \left [
\frac{a^2}{4} \sum_{x,i} \tr (\dot U^{\dagger}_{x,i} \dot U_{x,i}) +
\sum_{x,ij} (1-\oh \tr U_{x,ij}) \right ],
\ee
where $U_{x,ij}$ again stands for a plaquette quaternion.
The second term in the Hamiltonian descibes the free Higgs doublet
and its coupling to the gauge field
\be
H_{\rm HW} = {\beta_{\rm H}\over a} \left [
\sum_x \frac{a^{2}}{2} \tr \dot \Phi^{\dagger}_x \dot \Phi_x +
\sum_{x,i} \oh \tr \Phi^{\dagger}_x (\one - U_{x,i}) \Phi_{x+i} \right ]
\ee
where Higgs fields on neighboring sites $x$ and $x+i$ are
coupled by the lattice transport operator $(\one - U_{x,i})$,
that generates the gauge covariant derivative in the continuum limit.
The last term in the Hamiltonian accounts for the self-interaction
of the Higgs fields
\be
H_{\rm H} = {\beta_{\rm R}\over a} \sum_x (R_x - 1)^2,
\ee
where $R_x$ is defined by eq. (\ref{R}).
The coupling parameters $\beta_{\rm G}$, $\beta_{\rm H}$, and
$\beta_{\rm R}$, introduced into the lattice Hamiltonian are related
to the parameters of the original continuum Hamiltonian as follows
\be
\mu^2 = \frac{\beta_{\rm R}}{\beta_{\rm H}} \frac{1}{a^2},
\qquad
\lambda = \frac{\beta_{\rm R}}{\beta_{\rm H}^2},
\qquad
g^2 = \frac{4}{\beta_{\rm G}}.
\ee
The vacuum expectation value of the Higgs field in
the spontaneously broken phase is obtained from the minimum
of the energy term $H_{\rm H}$, and can be scaled to unity.
The symmetry breaking, where $\avrg {R_x} =1$,
occurs when
\be
\beta_{\rm H} > \frac{\hbar}{3} - \frac{\hbar^2}{27\beta_{\rm G}}.
\ee
In this phase large Higgs and gauge boson masses are generated:
\be
M^2_{\rm H} = 4 \hbar^2 \lambda \avrg{\Phi^2}
= \frac{4 \beta_{\rm R}}{\beta_{\rm H}} \frac{\hbar^2}{a^2}
\ee
and
\be
M^2_{\rm W} = \oh \hbar^2 g^2 \avrg{\Phi^2}
= \frac{2\beta_{\rm H}}{\beta_{\rm G}} \frac{\hbar^2}{a^2}.
\ee
The equations of motion derived from the lattice Hamiltonian
\be
H = H_{\rm W} + H_{\rm HW} + H_{\rm H},
\ee
are identical to (\ref{U0dot}, \ref{Udot}) for the time derivatives
of the components of the link variables, $u_{x,i}^a$ and $u_{x,i}^0$.
The three electric fields $E^c_{x,i}$ on each link are updated according to
\be
\dot E^a_{x,i} = \frac{i}{a g^2} \sum_{j}
\tr [\oh \tau^a (U_{x,ij} - U^{\dagger}_{x,ij})]
+ \frac{\beta_{\rm H}}{\beta_{\rm G}} \tr (\oh \tau^a U_{x,i}
\Phi_{x+i} \Phi^{\dagger}_x),
\label{EdotH}
\ee
where the sum again runs over all four plaquettes
attached to the link $(x,i)$. Gauss' law now has the form
\be
D^{ab}_i E^b_i - i g (\Phi^{\dagger} \tau^a \dot \Phi -
\dot \Phi^{\dagger} \tau^a \Phi)  = 0,
\label{GaussH}
\ee
and remains conserved under time evolution.
Finally, the Higgs fields evolve according to
\be
\ddot \Phi_x = \sum_i (U_{x,i} \Phi_{x+i} + U_{x,-i} \Phi_{x-i})
- [6+\frac{4 \beta_{\rm R}}{\beta_{\rm H}} (R^2_x-1)] \Phi_x,
\label{Phiddot}
\ee
where the summation index $i$ runs over all spatial directions
$(i=x,y,z)$. By introducing an auxiliary variable $\Psi_x = \dot\Phi_x$, 
we transform the set of second-order differential equations
into a larger set of coupled first-order differential equations.

Solving the evolution equations (\ref{Udot},\ref{EdotH},\ref{Phiddot})
we can study the dynamics of the coupled Yang-Mills-Higgs system,
and see how the fields approach thermal equilibrium.
In particular, we can investigate the divergence of trajectories
in the configuration space. In addition to the already defined
gauge-invariant distance measure (\ref{eq10})  for the distance between two
Yang-Mills fields which is based on the magnetic energy,
we measure the distance of Higgs field configurations by
\be
D^H[\Phi, \Phi'] = \frac{1}{N^3} \sum_x |R_x - R'_x|,
\ee
which is also gauge invariant.

Depending on the relative strength of the gauge coupling $g$
and Higgs self-coupling $\lambda$ different hierarchies
in the thermalization rate can be obtained. Figure 16 shows
the time evolution of the gauge field and Higgs-field distances
(a) in a cross-coupled, $\lambda \approx g$ and (b) in
a self-coupling dominated case, $\lambda \gg g$.
The logarithmic growth rate, defined as
\be
h = \frac{d}{dt} \ln D(t),
\ee
again depends on the average energy per plaquette.
This dependence, counting only the energy per plaquette contained in the
gauge field, is somewhat different in the above two cases, 
$\lambda \approx g$ and $\lambda \gg g$, as shown in Figure 17. 
At the upper end of the energy scale, $h(E)$ is about the same for
the two cases and also agrees with the value (\ref{eq14}) for the
pure gauge field, $h(E)$ is much smaller for the strongly coupled
case (b) for low energies. We also note that the gauge 
field becomes chaotic faster than the Higgs field in case (b), while 
they become chaotic equally fast in case (a).

\section{Thermalization of Gauge Fields}

\subsection{Thermalization Time}

The universal exponential divergence of neighboring gauge field
configurations for the gauge groups SU(2) and SU(3) implies that the
entropy $S$ of an ensemble of gauge fields grows linearly with time
\cite{LL83}:
\be
S(t) = S_0 + t \sum_{\lambda_i>0} \lambda_i, \label{eq16}
\ee
until the available microcanonical phase space is filled and the
system is equilibrated. In perturbation theory, the average
energy $E$ per plaquette is related to the temperature as \cite{AAPS92}:
\be
E \approx {2\over 3} (n^2-1) T \qquad {\rm [for}\;{\rm SU}(n){\rm ]}. 
\label{eq17}
\ee
It follows that the characteristic entropy growth rate, 
i.e. the thermalization rate is given by
\be
\Gamma_0\equiv \lambda_0 \approx \cases{0.34\; g^2T &[SU(2)] \cr
0.54\; g^2T &[SU(3)] \cr}, \label{eq18}
\ee
where we have inserted our best-fit numerical values from eqs. (\ref{eq14})
and (\ref{eq15}). Apart from a factor two, these values are remarkably
close to those for the thermal damping rate for a gauge boson at rest 
obtained by Braaten and Pisarski \cite{BP90}
\be
\gamma_0 = 6.635\; {N\over 24\pi} g^2T = \cases{ 0.175\; g^2T &[SU(2)], \cr
0.264\; g^2T &[SU(3)].\cr} \label{eq19}
\ee
At first sight, the relation $\Gamma_0 \approx 2\gamma_0$ is quite
surprising because these two quantities
appear in totally different contexts and are calculated
with different methods. On the one hand, the damping rate is
the imaginary part of the self energy of a quasi-particle
in a thermal gauge system and is calculated in the framework of
effective quantum field theory. On the other hand, the Lyapunov
exponent is a classical dynamical quantity describing the divergence 
of two classical gauge field trajectories.

Although we do not yet know how to establish a direct relation between 
these two quantities, we understand that this similarity does not
arise without reason. The two quantities, though very different from
their contexts, both describe how fast a non-equilibrated
gluon system approaches thermal equilibrium.
The relevance of $\lambda_0$ is clear from the above discussion.
The connection of $\gamma(\omega)$ is apparent from the relation
\cite{We82}
\begin{equation}
f(\omega,t)=\frac{1}{e^{\omega /T} - 1}
+c(\omega)e^{-2\gamma (\omega) t},
\end{equation}
where $f(\omega,t)$ is time-dependent gauge boson distribution function,
and the first term on the right-hand side is the Bose distribution. 
Obviously, $2\gamma$ describes the rate of approach to equilibrium.
It is possible to show that the gluon damping rate is basically
a quantity of semi-classical origin \cite{Go93}.

Finally, let us estimate the gluon thermalization quantitatively.
Figure 18 shows $\tau_S=\Gamma_0^{-1}$ as function of temperature. 
For the gauge coupling constant $g^2$ we have used the renormalized
running coupling constant of SU(3) gauge theory, which in the one-loop
approximation is given by:
\be
g^{2}(T) = \frac{16\pi^{2}}{11 \ln (\pi T/\Lambda)^{2}},
\ee
where $\Lambda\approx$ 200 MeV is the QCD scale parameter.
For temperatures in the range $T=300-500$ MeV, which are realistically
accessible in relativistic heavy-ion collisions, the thermalization
time is less than 0.4 fm/$c$, or about $10^{-24}$ s. This
promises the rapid formation of a locally thermalized gluon
plasma in these collisions at sufficiently high energy.

\subsection{Self-Thermalization of Gauge Fields}

With the numerical tools in this context we can study the thermalization
of gauge fields by direct simulation of their real-time evolution. 
Here we show that a nonabelian gauge field far off equilibrium approaches 
a thermally equilibrated state very rapidly.
A thermally equilibrated state is the state in which the
energy distribution over the microscopic degrees of freedom does not
change with time and takes the thermal equilibrium form, which can be 
calculated from the canonical Gibbs ensemble.
In order to study the process of thermalization,
we must decide on which particular energy distribution we want
to monitor.  Here we choose as our monitor $P(E_{M})$, the distribution
of magnetic energy on the elementary plaquettes over the whole
lattice. The reason we prefer magnetic energy to
electric energy is that the thermal distribution of the former
can also be obtained using a heat bath algorithm, and so
we can compare the results from time evolution and those from
a heat bath method. $P(E_{M})$ is obtained by counting the number of
plaquettes that have magnetic energy between $E_{M}-\Delta E_{M}/2$
and $E_{M}+\Delta E_{M}/2$.
$\Delta E_{M}$ should be sufficiently small so that $P(\Em)$ is smooth 
but large enough to provide good counting statistics.
Starting from an arbitrary intial state, we can measure the time
evolution of $P(E_{M})$. If we find that it reaches some stable
form we conclude that the system is thermally equilibrated.

We numerically integrate the equations of motion for given
initial conditions and measure the energy distribution functions $P(\Em,t)$.
The behavior in SU(2) and in SU(3) is rather similar. 
Here we concentrate on the SU(3) results. In Figure 19 we show the time
evolution of $P(\Em)$ for SU(3). The corresponding
initial averaged energy per plaquette is $E=\langle E_M\rangle=1.73$.
The plot shows
$\ln(P(\Em,t)/\Em^{3})$ for reasons that will become clear below.
The solid line denotes the initial distribution at $t=0$.
The dotted and short-dashed lines are for $t=0.5$
and $t=1.5$ respectively. The long-dashed line shows the
final distribution reached at $t=3$, whereafter no noticeable
change is observed.
The thermalization time is compatible with the inverse of
the maximal Lyapunov exponent, which is 
$\lambda_0^{-1}=10 (g^2 E)^{-1}=7.7$
for SU(3). The final magnetic energy per plaquette is 0.84 which is almost
half of the total energy.  This is not always 
true on the lattice where the magnetic energy is limited
from above because of the compact nature of the gauge group manifold.
But in our example the total energy is small and the compactness is not 
a relevant influence.

We have obtained equilibrated distributions $P(\Em)$
for different initial energies in both SU(2) and SU(3).
These distributions can all be almost perfectly fit by
\BE
P(\Em)={\cal N}f(\Em)\exp(-\Em/T_{s}),
\EE
where ${\cal N}$ is a normalization constant,
and $T_{s}$ is a ``temperature'' parameter.
$f(E)$ is the single plaquette phase space factor defined as
\BE
f(\Em)d\Em = \int \delta(\Em-E[U]) d\Em d\mu(U),
\EE
where  $E[U]=2\re(n-\tr U)$
is the magnetic energy of a single free plaquette
for SU($n$) and $d\mu(U)$ is the Haar measure.  Although the 
distribution looks like a Boltzmann distribution and $T_{s}$ like a
temperature, we have shown that the real temperature of the system differs
from $T_{s}$. Their relation is shown in Figure 20 for SU(2). 
While the ratio $T_{s}/T\to 1$ at high temperature,
it decreases with $T$ when $T<2$ and finally it
reaches ${2\over 3}$ in the limit $T=0$.
We expect a similar behaviour for SU(3).
This implies that in the low temperature limit the number of effective
degrees of freedom is 2/3 of that in the high temperature limit.

\section{Conclusion and Future Developments}

The studies of lattice gauge fields reviewed here have considerably
extended our knowledge of the properties of the dynamics of classical
nonabelian gauge fields. This work began with the results of Matinyan,
Savvidy and others more than a decade ago. We now know that the
classical Yang-Mills fields are strongly chaotic at all energies.
We have obtained the complete Lyapunov spectrum and shown that it
scales even for very small lattice sizes. We are thus close to a full
understanding of the classical Yang-Mills equations from the point
of view of classical dynamics. We have also learned that, although
other field theories show signs of chaotic behavior on the lattice, 
as well, their scaling properties imply that the chaoticity disappears
in the continuum limit. This suggests that the chaotic behavior of
nonabelian gauge theories is a nontrivial property of these theories.

What are the consequences of the chaotic nature of the 
classical Yang-Mills theory? We have already 
studied its application to thermalization at finite temperature.
Another possible application is the old idea \cite{Ol82}
that the chaoticity may be related to the problem of color confinement. 
In order to investigate this conjecture,
as well as other implications, more thoroughly, the following 
two possible directions of future research are important.

In order to connect the results of classical gauge field dynamics
to the underlying quantum field theory,
we must understand the corrections from quantum effects. 
This means that we must develop appropriate
semi-classical methods that link the classical limit to
quantum physics. We have recently begun to explore a variational
method using Gaussian wave packets for lattice gauge fields 
and found indications that the quantum corrections
enhance the chaoticity \cite{GMB93}. However, it is known that when a
system is chaotic in the classical limit, the quantum evolution
of a initially localized Gaussian wave packet tends to
disperse the wave packet rapidly \cite{FMR91}. This is easily understood
in the path integral formalulation of the quantum theory.
A more realistic approach must account for this property. 
One possible approach is the Wigner function formulation mentioned
in section 2.1. However, Wigner functions for gauge fields involve 
many subtleties
\cite{EGV86}, and it is not quite clear how the definition of a Wigner
function should be best applied to gauge fields on a lattice.

Secondly, the studies have so far been restricted to gauge fields and 
scalar fields.  Since, in reality, the gauge fields are always coupled 
to fermion fields, a natural question is what role the fermions
play in this context. Fermions cannot be treated classically.
As a possible solution, on could consider a hybrid model decribing the 
interaction of the quantum evolution of matter fields with
the classical evolution of the gauge fields. This requires the
solution of the time-dependent Dirac equation on the lattice, 
which is equivalent to a quantum cellular automaton
\cite{Bi93}.

We must bear in mind that numerical studies of the real-time
evolution of (quantum) field theories on lattices are in their
infancy. The results obtained so far indicate that interesting results 
can be obtained from such investigations, and there is reason to hope
that further progress is possible.

\bigskip

{\it Acknowledgment:} This work has been supported in part by
the U. S. Department of Energy (Grant DE-FG05-90ER40592) and by
a computing grant from the North Carolina Supercomputing Center.
We thank S. Matinyan for many illuminating discussions. One of us
(B.M.) thanks the Institute of Nuclear Theory at the University
of Washington for its hospitality and the Department of Energy
for partial support during the completion of this work.

\eject



\begin{figure}
\centerline{\includegraphics[width=0.5\textwidth]{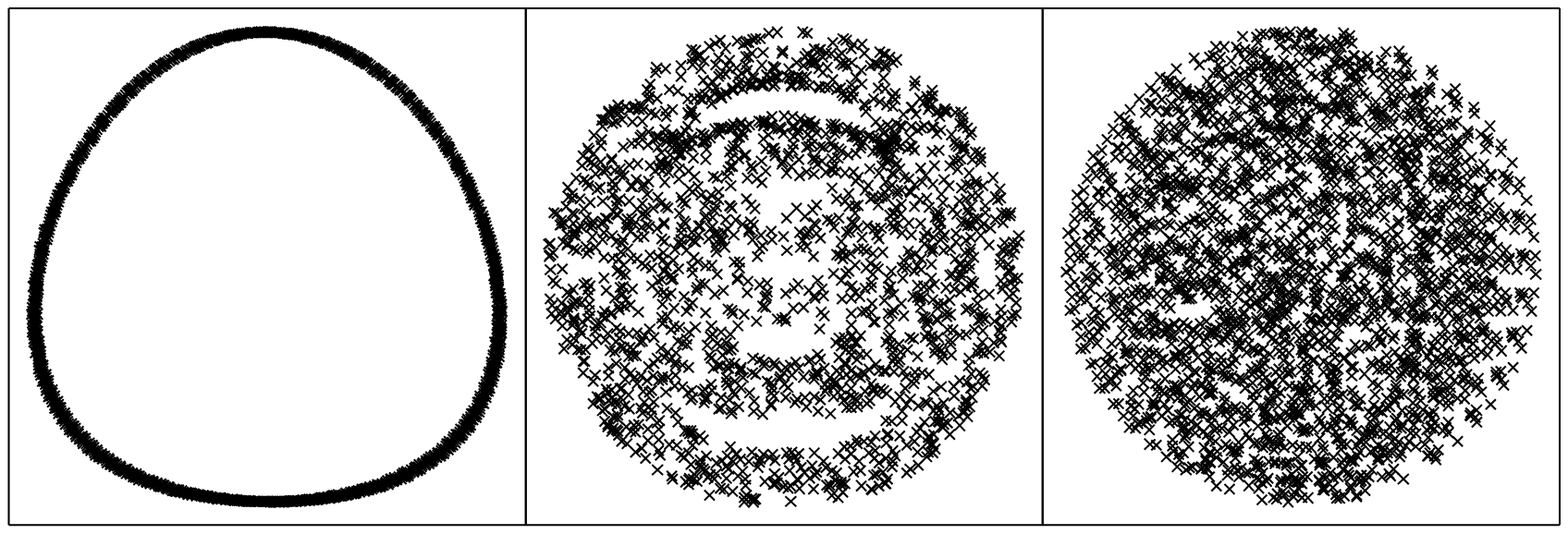}}
\caption{
Poincar\'e surfaces of section of a single trajectory in the
two-dimensional Yang-Mills model $(z(t)\;\equiv\; 0)$ for
three values of the scaling parameter: $r=4.878$ (left), $r=0.2$ (center),
and $r=0.0012$ (right).}
\end{figure}\newpage

\begin{figure}
\centerline{\includegraphics[width=0.5\textwidth]{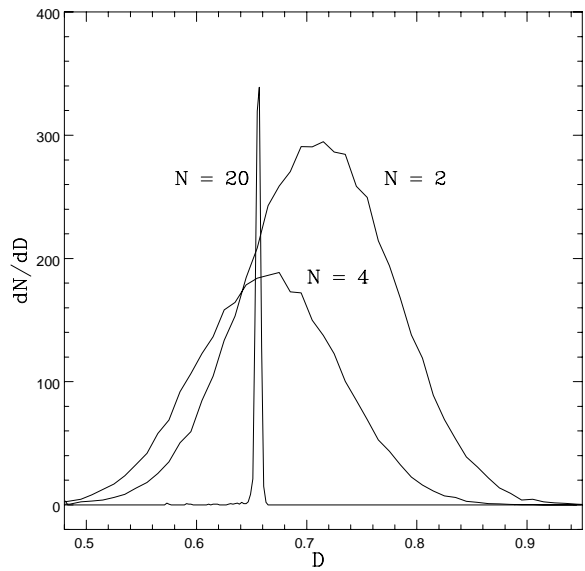}}
\caption{
Distance distribution of points with the same distance $D$ from a given 
SU(2) field configuration for lattices of size $2^3, 4^3$, and $20^3$.}
\end{figure}\newpage

\begin{figure}
\centerline{\includegraphics[width=0.5\textwidth]{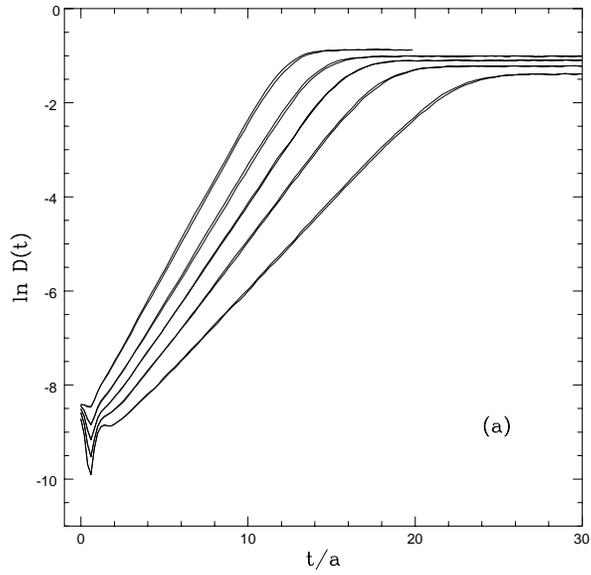}}
\centerline{\includegraphics[width=0.5\textwidth]{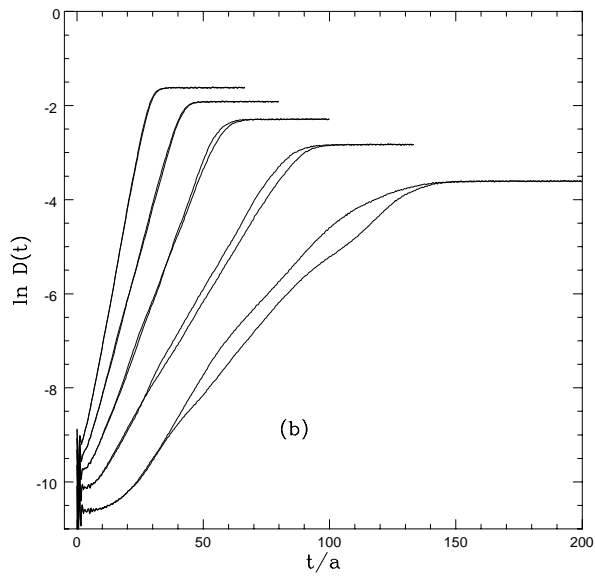}}
\caption{
Evolution of the distance $D(t)$ between neighboring random gauge
field configurations for several average energies
on a $20^3$ lattice.  The curves correspond, from top to
bottom, to the parameters (a) $\delta = 1, 0.5, 0.45, 0.4, 0.35$;
(b) $\delta = 0.3, 0.25, 0.2, 0.15, 0.1$.
For every value of $\delta$ two curves are shown, which are
indistinguishable when $\delta > 0.2$.}
\end{figure}\newpage

\begin{figure}
\centerline{\includegraphics[width=0.5\textwidth]{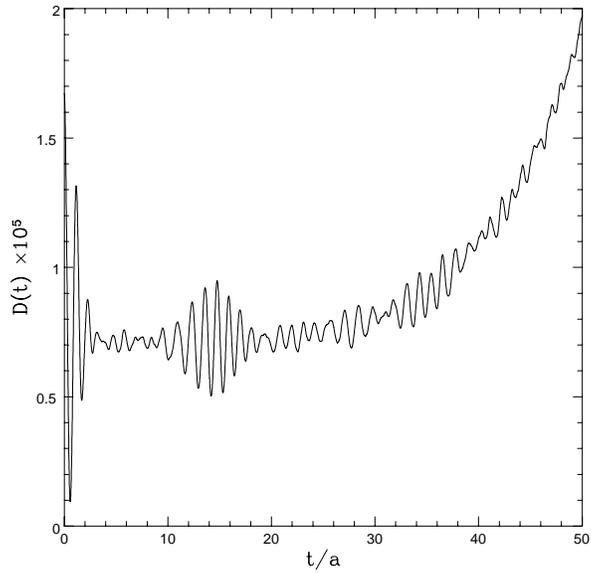}}
\caption{
Initial oscillations in the distance between neighboring field
configurations, before the exponential divergence sets in (for
$\delta=0.03$ and $N=10$).}
\end{figure}\newpage

\begin{figure}
\centerline{\includegraphics[width=0.5\textwidth]{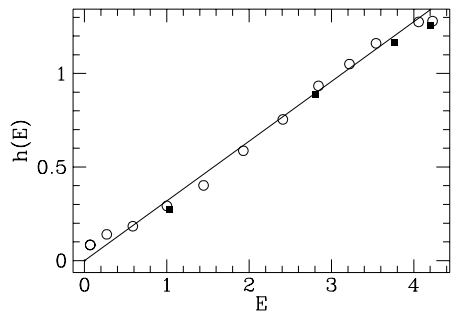}}
\caption{
Dependence of the exponential growth rate $h$ on
the average energy per plaquette $E$ of the randomly chosen
field configuration, for $a=0.5$ and $4/g^4 =
1.1185$, and for lattice sizes $N=20$ (open circles) and
$N=6$ (solid squares). The straight line through the
origin is a least-squares fit.}
\end{figure}\newpage

\begin{figure}
\centerline{\includegraphics[width=0.5\textwidth]{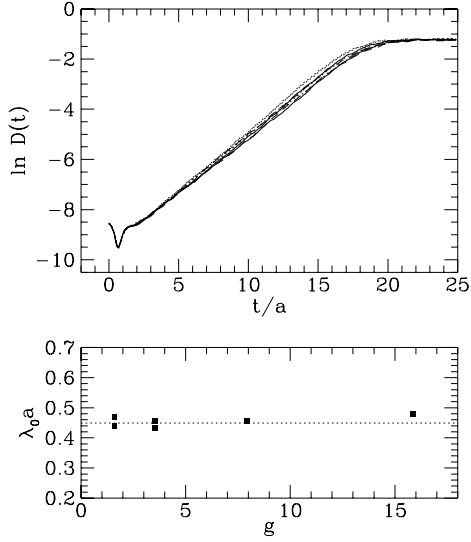}}
\caption{
Scaling of the maximal Lyapunov exponent $\lambda_0$ with the variable
$g^2Ea$. At fixed value of the scaling variable, the exponent is independent 
of the value of $g^2$ over a wide range. The top part shows $D(t)$ for
different choices of $g^2$, the bottom part shows the Lyapunov exponent.}
\end{figure}\newpage

\begin{figure}
\centerline{\includegraphics[width=0.5\textwidth]{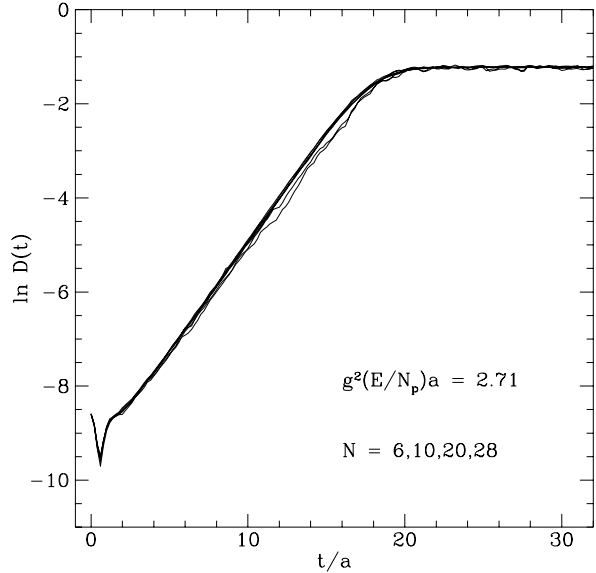}}
\caption{
Dependence of the evolution of the
distance between field configurations on the lattice size $N$.  
The curves for $N=6,\; 10,\; 20$, and 28 nearly coincide.  
All curves correspond to $\delta = 0.4$.}
\end{figure}\newpage

\begin{figure}
\centerline{\includegraphics[width=0.5\textwidth]{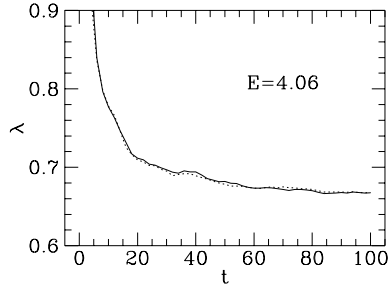}}
\caption{
The two largest Lyapunov exponents for SU(2) determined by the rescaling
method. The average of the logarithmic scaling factors $s_i^{k}$
approaches the limit from above.}
\end{figure}\newpage

\begin{figure}
\centerline{\includegraphics[width=0.5\textwidth]{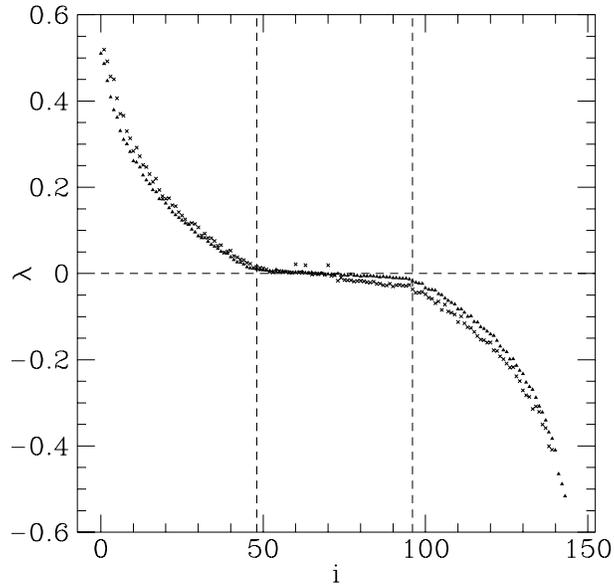}}
\caption{
Complete spectrum of 144 Lyapunov exponents for SU(2) gauge theory 
on a $2^3$ lattice. The trajectories were followed up to time $t/a=200$
(crosses) and $t/a=1000$ (triangles). The central third fraction of 
Lyapunov exponents (enclosed between the vertical dashed lines)
corresponds to the unphysical degrees of freedom that describe gauge 
transformations and deviations from Gauss' law. These exponents 
converge to zero in the limit $t\to\infty$.}
\end{figure}\newpage

\begin{figure}
\centerline{\includegraphics[width=0.5\textwidth]{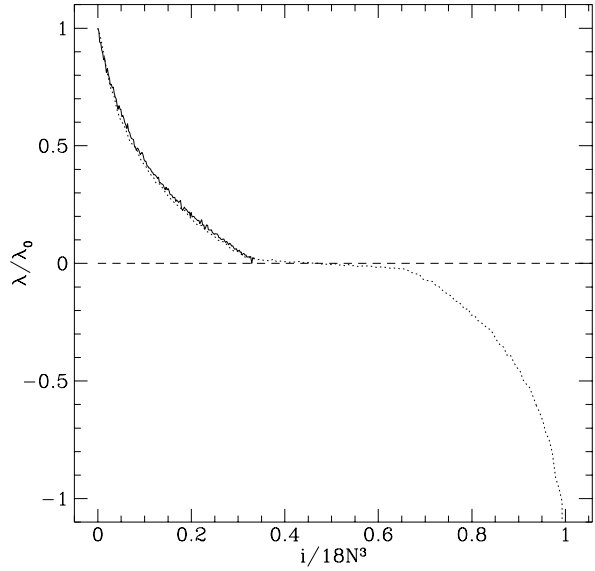}}
\caption{
Scaling of the Lyapunov spectrum with lattice size $N$. The solid line
corresponds to a $3^3$ lattice; the dashed line is for a $2^3$ lattice.
Only the positive Lyapunov exponents are shown.
The exponents $\lambda_i$ are scaled with the maximal Lyapunov exponent 
for each lattice size, and the index $i$ is scaled with $N^3$.}
\end{figure}\newpage

\begin{figure}
\centerline{\includegraphics[width=0.5\textwidth]{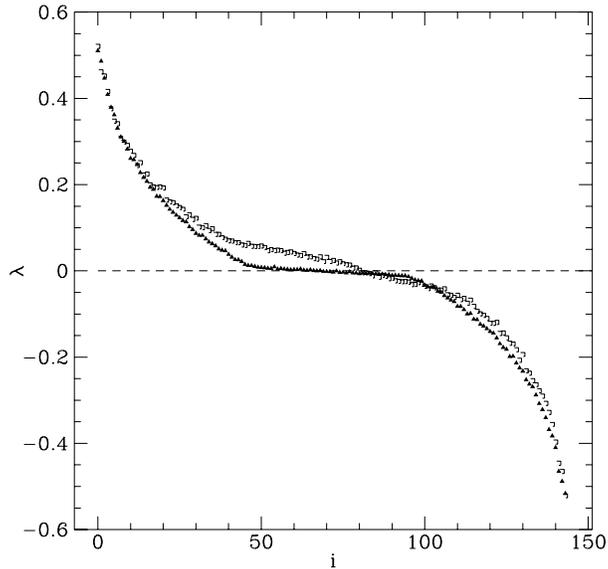}}
\caption{
Lyapunov spectrum for a $2^3$ lattice obtained with the second
scaling method working directly in the tangent space (hollow square),
in comparison with the results based on the first method (solid triangle).}
\end{figure}\newpage

\begin{figure}
\centerline{\includegraphics[width=0.5\textwidth]{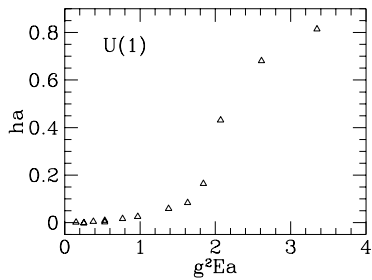}}
\caption{
Dependence of the scaled growth rate $ha$ on the 
scaling parameter $g^2Ea$ for the gauge group U(1) on a
$10^3$ lattice.  Note the highly nonlinear behavior and
the rapid vanishing of $ha$ in the limit $g^2Ea\to 0$.}
\end{figure}\newpage

\begin{figure}
\centerline{\includegraphics[width=0.5\textwidth]{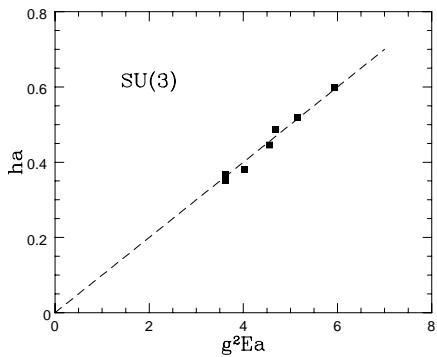}}
\caption{
Dependence of the scaled exponential growth rate $ha=\lambda_0 a$ 
on the scaling parameter $g^2Ea$ for the gauge group SU(3).
The calculations were performed on a $10^3$ lattice.  The
dashed line depicts the fit by eq. (39).}
\end{figure}\newpage

\begin{figure}
\centerline{\includegraphics[width=0.5\textwidth]{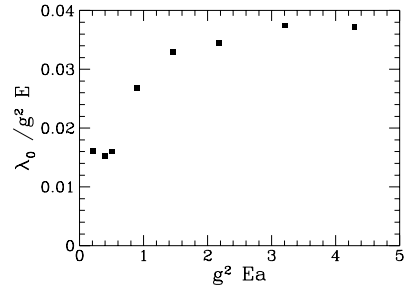}}
\caption{
Maximal Lyapunov exponent for the massless $\Phi^4$ theory as function
of the scaling variable $g^2Ea$.}
\end{figure}\newpage

\begin{figure}
\centerline{\includegraphics[width=0.5\textwidth]{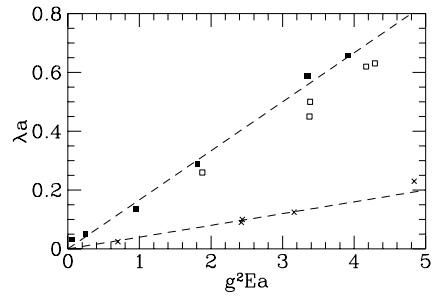}}
\caption{
Maximal Lyapunov exponent for the massive vector field theory as function
of the scaling variable $g^2Ea$ for two different values of the vector
boson mass: $m=0.2$ (hollow squares), $m=4$ (crosses), and for comparison
for $m=0$ (solid squares).}
\end{figure}\newpage

\begin{figure}
\centerline{\includegraphics[width=0.5\textwidth]{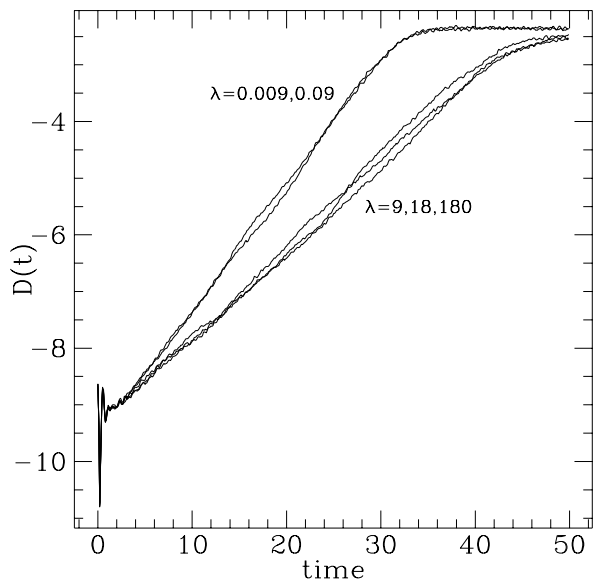}}
\caption{
Time evolution of the gauge field distance and the Higgs field
distance for the spontaneously broken SU(2) gauge theory. Part (a)
shows a case where the gauge coupling $g$ and the Higgs self-coupling
$\lambda$ are about equal; part (b) corresponds to the case $\lambda\gg g$.
For the strongly coupled Higgs field (case b) most of the chaoticity
resides in the gauge field.}
\end{figure}\newpage

\begin{figure}
\centerline{\includegraphics[width=0.5\textwidth]{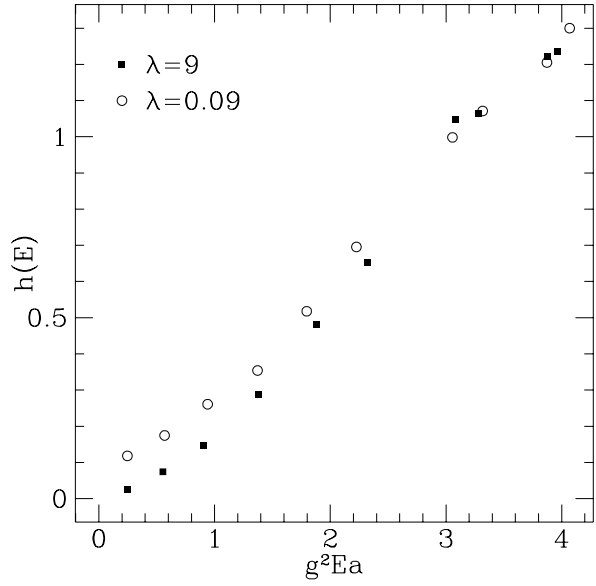}}
\caption{
Maximal Lyapunov exponent of the Yang-Mills Higgs field as function
of the variable $g^2Ea$ for the two cases discussed in the caption
of Figure 16: (a) $\lambda=0.9$, (b) $\lambda=9$. The gauge coupling
was $g=1.375$, and the lattice spacing $a=0.5$. The weakly coupled Higgs 
field follows very closely the results obtained for the pure gauge field.}
\end{figure}\newpage


\begin{figure}
\centerline{\includegraphics[width=0.5\textwidth]{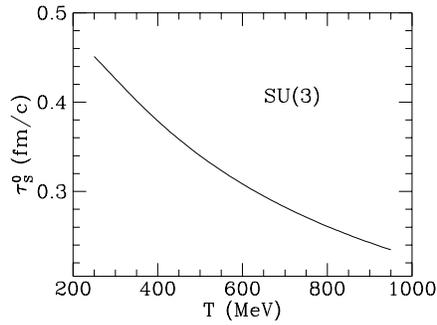}}
\caption{
``Thermalization time'' $\tau_S^0=\lambda_0^{-1}$ for SU(3) as function of 
temperature $T$. The scale parameter was taken as $\Lambda$ = 200 MeV.}
\end{figure}\newpage

\begin{figure}
\centerline{\includegraphics[width=0.5\textwidth]{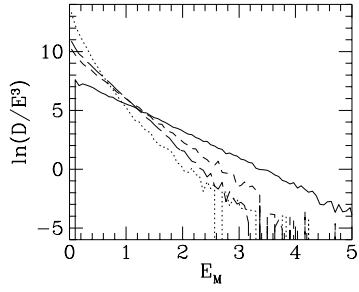}}
\caption{
Time evolution of the distribution of magnetic plaquette energies $\Em$
for a SU(3) gauge field configuration. The plot shows the quantity
$P(\Em)/\Em^3$ on a logarithmic scale. The change in slope from the
initial distribution (solid line) to the final distribution (long-dashed
line) corresponds to the thermalization of all electric field modes,
which has a cooling effect on the initially populated magnetic modes.}
\end{figure}\newpage

\begin{figure}
\centerline{\includegraphics[width=0.5\textwidth]{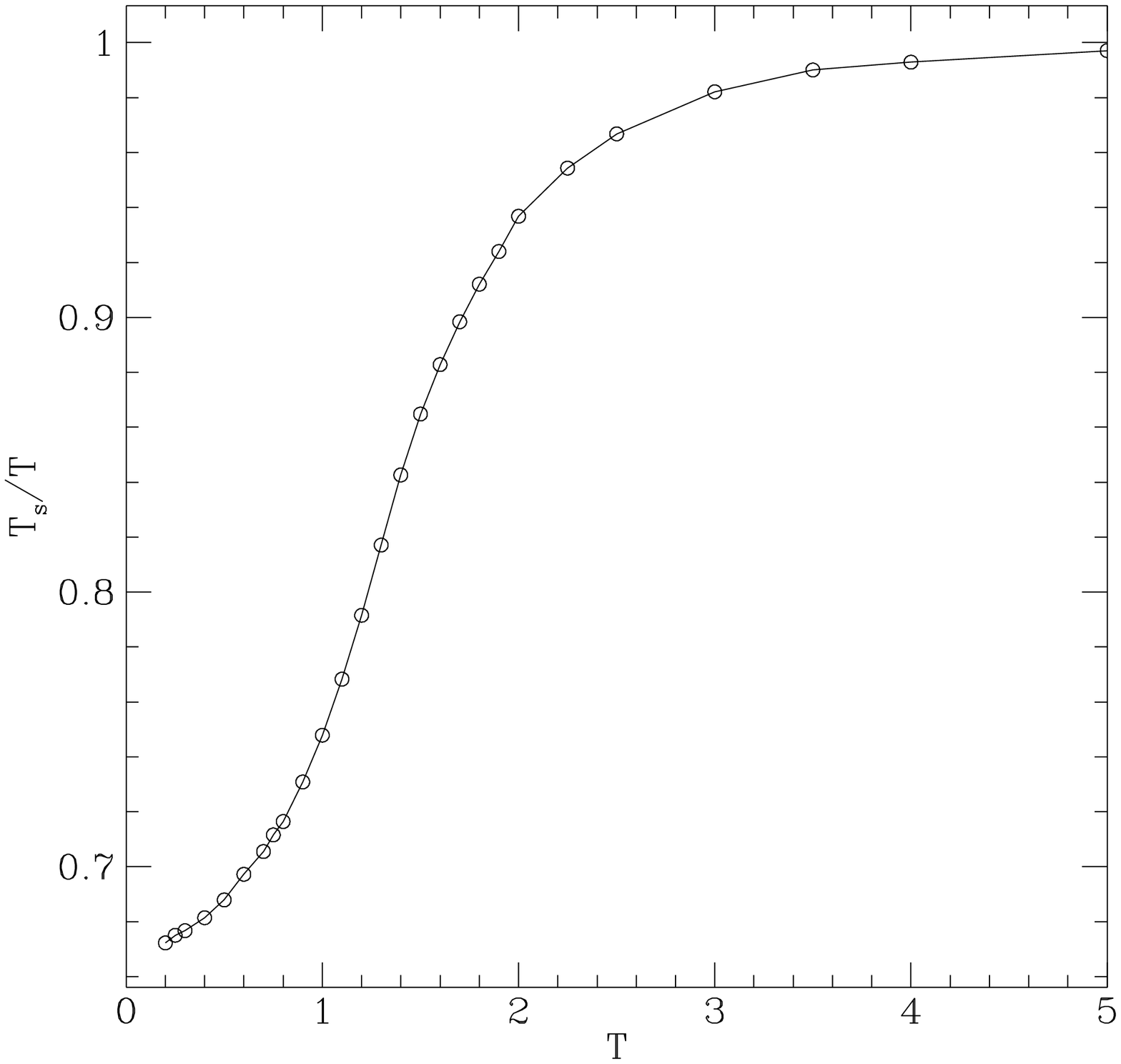}}
\caption{
Slope parameter (``apparent temperature'') $T_s$ of the magnetic
energy distribution $P(\Em)$ for the SU(2) gauge field as function
of real temperature $T$. The plot shows the ratio $T_s/T$, which
rises from ${2\over 3}$ to 1, because of contributions from the
longitudinal plasma modes at higher temperature.}
\end{figure}\newpage


\begin{thebibliography}{99}

\bibitem{}
	{\it Electronic mail addresses:} \\
	T.S. Bir\'o: ug12@ddagsi3.gsi.de,
	C. Gong: gong@phy.duke.edu,\\
	B. M\"uller: muller@phy.duke.edu,
	A. Trayanov: nasco@phy.duke.edu.

\bibitem{FM92}
	see e.g.: J. Ford and G. Mantico, Amer. J. Phys. {\bf 60}, 1086
	(1992).

\bibitem{FMR91}
	J. Ford, G. Mantico, and G.H. Ristow, Physica {\bf D50}, 493
	(1991).

\bibitem{Wi32} 
	E. Wigner, Phys. Rev. {\bf 40}, 749(1932) .

\bibitem{JR87} 
 	S. John and E. A. Remler, Ann. Phys.
        {\bf 180}, 152 (1987),  and refernces therein.

\bibitem{EH90} 
	H.T. Elze and U. Heinz, Phys. Rep. {\bf 183}, 81 (1989).

\bibitem{Wi74}
	K. Wilson, Phys. Rev. {\bf D10}, 2445 (1974).

\bibitem{PY57}
	R.E. Peierls and J. Yoccoz, Proc. Phys. Soc. (London) {\bf A70},
	381 (1957).

\bibitem{HP74}
	G. 't Hooft, Nucl. Phys. {\bf B79}, 276 (1974);
	A.M. Polyakov, JETP Lett. {\bf 20}, 194 (1974).

\bibitem{BPST75}
	A.A. Belavin, A.M. Polyakov, A.S. Schwartz, and Yu.S. Tyupkin,
	Phys. Lett. {\bf 59B}, 85 (1975).

\bibitem{MST81}
	S. G. Matinyan, G. K. Savvidy and N. G. Ter-Arutyun\-yan-Savvidy,
	Sov. Phys. JETP {\bf 53}, 421 (1981);
	JETP Lett. {\bf 34}, 590 (1981).

\bibitem{BP90}
	E. Braaten and R. D. Pisarski, Phys. Rev. {\bf D42}, 2156 (1990).

\bibitem{BLS81} 
	A. Billoire, G. Lazarides, and Q. Shafi, Phys. Lett.
	{\bf 103B}, 450 (1981);
	T.A. DeGrand and D. Toussaint, Phys. Rev. {\bf D25}, 526 (1981).

\bibitem{BM93} 
	T.S. Bir\'o and B. M\"uller, Nucl. Phys. {\bf A} (in print).

\bibitem{CS81}
	B. V. Chirikov and D. L. Shepelyanskii, JETP Lett. {\bf 34},
	163 (1981); Sov. J. Nucl. Phys. {\bf 36}, 908 (1982).

\bibitem{NS82}
	E. S. Nikolaevskii and L. N. Shchur, JETP Lett.
	{\bf 36}, 218 (1982); Sov. Phys. JETP {\bf 58}, 1 (1983).

\bibitem{Fr83}
	J. Fr\o yland, Phys. Rev. {\bf D27}, 943 (1983).

\bibitem{Sa83}
	G. K. Savvidy, Phys. Lett. {\bf 130 B}, 303 (1983).

\bibitem{Lu83}
	M. L\"uscher, Nucl. Phys. {\bf B19}, 233 (1983).

\bibitem{SLLM86}
	W. M. Steeb, J. A. Louw, P. G. L. Leach, and
	F. M. Mahomed, Phys. Rev. {\bf A33}, 2131 (1986).

\bibitem{DR90}
	P. Dahlqvist and G. Russberg,
	Phys. Rev. Lett. {\bf 65}, 2837 (1990).

\bibitem{MPS88}
	S. G. Matinyan, E. B. Prokhorenko and G. K. Savvidy,
	Nucl. Phys. {\bf B298}, 414 (1988).

\bibitem{KO90}
	T. Kawabe and S. Ohta, Phys. Rev. {\bf D41}, 1983 (1990).

\bibitem{JS89}
	M. P. Joy and M. Sabir, J. Phys. {\bf A22}, 5153 (1989).

\bibitem{We92}
	M. Wellner, Phys. Rev. Lett. {\bf 68}, 1811 (1992).

\bibitem{MT92}
	B. M\"uller and A. Trayanov, Phys. Rev. Lett.
	{\bf 68}, 3387 (1992).

\bibitem{Cr83}
	M. Creutz, {\sl Quarks, Gluons, and Lattices}
	(Cambridge Univ. Press, Cambridge, 1983).

\bibitem{KS75}
	J. Kogut and L. Susskind, Phys. Rev. {\bf D11}, 395 (1975).

\bibitem{CRUK85}
	S. A. Chin, O. S. van Roosmalen, E. A. Umland, and
	S. E. Koonin, Phys. Rev. {\bf D31}, 3201 (1985).

\bibitem{Go93}
  	C. Gong, Phys. Lett. {\bf B 298}, 257 (1993) .

\bibitem{Br88}
	J. P. Bronzan, Phys. Rev. {\bf D38}, 1994 (1988).

\bibitem{Pi81}
	E. Pietarinen, Nucl. Phys. {\bf B190}, 349 (1981).

\bibitem{LL83}
	See e.g.: A. J. Lichtenberg and M. A. Lieberman, {\sl Regular
	and Stochastic Motion} (Springer-Verlag, New York, 1983).

\bibitem{BFS79}see for example,
        G. Benettin, C. Froeschle, and J. P. Scheidecker,
        Phys. Rev. {\bf A19}, 2454 (1979).

\bibitem{AAPS92}
	J. Ambj\o rn, T. Aksgaard, H. Porter, and M. E.
	Shaposhnikov, Nucl. Phys. {\bf B353}, 346 (1992).

\bibitem{We82}
	A. Weldon, Phys. Rev. {\bf D28}, 2007 (1982).
        Nucl. Phys. {\bf A522}, 591 (1991).

\bibitem{Ol82}
	P. Olesen, Nucl. Phys. {\bf B200}, 381 (1982).

\bibitem{GMB93}
	C. Gong, B. M\"uller, and T. Bir\'o,
	Duke University preprint DUKE-TH-93-49.

\bibitem{EGV86} 
	H.T. Elze, M. Gyulassy, and D. Vasak,
	Phys. Lett. {\bf B177} 402 (1986).

\bibitem{Bi93}
	I. Bialynicki-Birula, Univ. Warsaw preprint,
	$\langle$Bulletin Board: hep-th@xxx.lanl.gov - 9304070$\rangle$.

\end{thebibliography}
\end{document}